\documentclass[english]{article}

\usepackage[colorlinks,bookmarksopen,bookmarksnumbered,allcolors=red]{hyperref}
\usepackage[dvipsnames]{xcolor}
\usepackage{graphicx}
\usepackage{geometry}
\usepackage{amsmath, amssymb, bm}
\usepackage{relsize}
\usepackage{authblk}
\usepackage{setspace}
\usepackage{dsfont}
\usepackage{listings}
\usepackage{float} 
\usepackage{indentfirst} 

\lstdefinestyle{Rstyle}{
    language=R,
    basicstyle=\ttfamily\footnotesize,
    keywordstyle=\color{blue},
    commentstyle=\color{OliveGreen},
    stringstyle=\color{red},
    showstringspaces=false,
    frame=single, 
    breaklines=true,
    morekeywords={library, data} 
}

\definecolor{bluegreen}{RGB}{46,141,131}
\definecolor{darkgreen}{RGB}{46,139,87}
\definecolor{darkred}{RGB}{219,7,61}
\definecolor{darkblue}{RGB}{0,0,137}

\title{\textbf{The DeepJoint algorithm: An innovative approach for studying the longitudinal evolution of quantitative mammographic density and its association with screen-detected breast cancer risk}}

\author[,1]{Manel Rakez\thanks{Corresponding author: manel.rakez@u-bordeaux.fr}}
\author[2]{Julien Guillaumin}
\author[2]{Aurelien Chick}
\author[3,4]{Gaelle Coureau}
\author[5]{Foucauld Chamming's}
\author[2]{Pierre Fillard}
\author[$\dagger$,3]{Brice Amadeo}
\author[,1]{Virginie Rondeau\thanks{Brice Amadeo and Virginie Rondeau contributed equally as co-last authors}}
\affil[1]{BIOSTAT team, Bordeaux Population Health U1219, Bordeaux University, ISPED, Bordeaux, France}
\affil[2]{Therapixel, Nice, France}
\affil[3]{EPICENE team, Bordeaux Population Health U1219, Bordeaux University, ISPED, Bordeaux, France}
\affil[4]{Department of Public Health, Bordeaux University Hospital, Bordeaux, France}
\affil[5]{Department of Radiology, Institut Bergonié, Comprehensive Cancer Centre, Bordeaux, France}

\date{}

\begin{document}

\maketitle

\begin{abstract}
High mammographic density is a well-known risk factor for breast cancer and reduces the sensitivity of mammography-based screening. While automated machine and deep learning-based methods provide more consistent and precise measurements compared to subjective Breast Imaging Reporting and Data System (BI-RADS) assessments, they often fail to account for the longitudinal evolution of density. Many of these methods assess mammographic density in a cross-sectional manner, overlooking correlations in repeated measures, irregular visit intervals, missing data, and informative dropouts. Joint models address these limitations by simultaneously modeling the relationship between longitudinal biomarkers and time-to-event outcomes. We introduce the DeepJoint algorithm, an open-source method combining deep learning-based mammographic density estimation with joint modeling to assess its longitudinal relationship with breast cancer risk. Our approach adequately analyzes processed mammograms from various manufacturers, estimating both dense area and percent density, two established risk factors for breast cancer. We utilize a joint model to explore their association with breast cancer risk and provide individualized risk predictions. Bayesian inference and the Monte Carlo consensus algorithm make the approach reliable for large screening datasets. By integrating deep learning with joint modeling, our new method provides a robust, comprehensive framework for evaluating breast cancer risk based on longitudinal density profiles. The complete pipeline is publicly available, promoting broader application and comparison with other methods. \newline
\end{abstract}

\section{Introduction}
\label{sec::intro}

Mammographic density reflects the composition of breast tissue on a mammogram. Density levels vary widely among women, can evolve over time, and are visually assessed and categorized using the Breast Imaging Reporting and Data System (BI-RADS) into four categories: "entirely fatty" (A), "scattered areas of fibroglandular density" (B), "heterogenously dense" (C), and "extremely dense" (D) breasts.\cite{spak_bi-rads_2017} High mammographic density, defined as classes C and D, is a well-established risk factor for breast cancer. It can also reduce mammography's sensitivity, as dense tissue can mask potentially malignant lesions.\cite{boyd_origins_2018, boyd_mammographic_2007, pisano_etta_d_diagnostic_2005} While supposedly straightforward, the BI-RADS-based visual assessment remains subjective and prone to inter-reader variability, possibly resulting in limited reproducibility.\cite{portnow_persistent_2022}

Recent advances in artificial intelligence and deep learning have significantly improved breast imaging.~\cite{michel_breast_2023, balkenende_application_2022} These methods now play a central role in automating segmentation, risk assessment, and cancer detection tasks. Regarding mammographic density estimation, deep learning allows for fully automated, quantitative assessments that are more consistent and precise than BI-RADS-based evaluations.\cite{zhang_opinions_2022} Quantitative density is often described using two metrics: dense area, which measures the absolute amount of dense tissue in cm², and percent density, which reflects the proportion of dense tissue relative to total breast area. Both measures are independently associated with breast cancer risk.\cite{baglietto_associations_2014, krishnan_mammographic_2017}

Despite these advances, most existing approaches estimate mammographic density at a single time point.\cite{gudhe_area-based_2022, dembrower_comparison_2020} Thus, the evolution of mammographic density over time is ignored even though longitudinal changes in density may carry important information about breast cancer risk.\cite{jiang_longitudinal_2023, tran_association_2022} Furthermore, current machine and deep learning tools are often not designed to handle repeated measurements, irregular visit intervals, or missing data, all common in longitudinal screening datasets.\cite{cascarano_machine_2023}

Joint modeling of longitudinal and time-to-event data offers a statistically rigorous framework to address these challenges. Jointly modeling the temporal evolution of longitudinal biomarkers and time-to-event outcomes allows for dynamic risk prediction that updates as new data become available. This framework accounts for within-subject correlation, irregular measurement intervals, and informative dropout.\cite{rizopoulos_joint_2012, krol_joint_2016} Several recent studies have applied joint models to study the link between mammographic density and breast cancer.\cite{armero_bayesian_2016, illipse_studying_2023} For instance, Armero et al. (2016)\cite{armero_bayesian_2016} proposed a Bayesian joint model using a latent variable representing BI-RADS density categories. While effective, this method has limitations, including reliance on the subjective BI-RADS measure, complexity in the interpretation due to the latent variable, and a narrow focus on the current true value of the biomarker as the link function. More recently, Illipse et al. (2023)\cite{illipse_studying_2023} applied a joint model to analyze the association between longitudinal dense area measurements and breast cancer risk by exploring multiple link functions. Their analysis was based on data from a population-based prospective screening cohort. Dense area was estimated using the STRATUS method,\cite{Eriksson_comprehensive_2018} which quantifies this metric from processed mammograms obtained from multiple vendors. In their approach, the biomarker was estimated from the mediolateral oblique (MLO) view of either the contralateral (non-lesion) breast in women with breast cancer or a randomly selected breast in women without breast cancer. This method contrasts with current practice, where mammographic density is assessed for both breasts at each time visit, using two standard views per breast: MLO and craniocaudal (CC). Besides, STRATUS's limited open-source access prevents testing its reproducibility and extension to other settings. Consequently, there is a need to develop a method that estimates precisely density metrics and accounts for their evolution at the visit level.

Several tools exist for automated mammographic density estimation. Commercial software like Volpara\cite{hartman_volumetric_2008} and Quantra\cite{regini_radiological_2014} use x-ray beam models to estimate volumetric density, but rely on metadata that are not always available. Their cost and proprietary nature also hinder their general use. Academic tools\cite{byng_quantitative_1994, lee_automated_2018, lehman_mammographic_2019} offer alternatives but often suffer from restricted access, such as the STRATUS method, or limited training datasets. Recently, MammoFL\cite{muthukrishnan_mammofl_2023} has advanced quantitative density assessment by providing an open-source deep learning-based tool, improving upon previous methods like Deep-LIBRA\cite{haji_maghsoudi_deep-libra_2021} and LIBRA.\cite{keller_estimation_2012} This segmentation model employs two successive modified UNets\cite{Ronneberger_UNet_2015} to delineate breast and dense tissue areas from a mammogram. However, its dependence on raw mammograms from a single manufacturer poses a barrier to real-world implementation. Indeed, breast screening programs can involve images from multiple manufacturers, and raw images are rarely archived in clinical practice.\cite{shu_deep_2021, markets_united_2023} Thus, a more adaptable approach is needed to work across processed images and different manufacturers.

In this work, we introduce the DeepJoint algorithm, an open-source method that combines deep learning for mammographic density estimation with joint modeling to explore the longitudinal relationship between density metrics and screen-detected breast cancer risk. Our approach is designed to operate on processed images from multiple manufacturers, making it applicable in diverse clinical settings. It provides quantitative estimates of dense area and percent density at each screening visit and uses a joint model to link their trajectories with breast cancer risk. This approach enables dynamic, personalized risk prediction that adapts to each woman's screening history. We implement Bayesian inference\cite{rizopoulos_jmbayes2_2023} with the Consensus Monte Carlo algorithm\cite{afonso_efficiently_2023} to scale up to large screening datasets and allow efficient computation. All code for the \href{https://github.com/manelrakez/deepjoint-algo}{DeepJoint algorithm} is publicly available, supporting transparency, reproducibility, and further development.

The remainder of this paper is structered as follows: Section~\ref{sec::DL_model} outlines details on the deep learning model for mammographic density estimation. Section~\ref{sec::JM_method} delineates the joint model for density association with breast cancer risk. Section~\ref{sec::results} presents our findings, and Section~\ref{sec::discuss} discusses the implications of our results, with some concluding remarks.
\section{The deep learning model for mammographic density estimation}
\label{sec::DL_model}

The deep learning component of the DeepJoint algorithm is embodied by a new fine-tuned version of the segmentation model MammoFL\cite{muthukrishnan_mammofl_2023} that works across processed images and different manufacturers.

\subsection{Image selection}

An in-house dataset was used in this work, referring to a dataset owned and maintained by Therapixel, a French AI-focused company.\cite{TPX_url} Data included retrospectively collected processed full-field digital mammography (FFDM) images from women screened between October 2006 and August 2019 in the United States. Images were acquired using Lorad Selenia and Selenia DimensionsTM units from Hologic Inc, and Senograph essential and Senograph DS units from General Electric Healthcare (GEHC). Our study is exempt from Subpart A of the federal regulations for the protection of human subjects 45 CFR 4 as it involves retrospective collected and de-identified data only, and it presents minimal risks.

We excluded duplicate images, images captured in views other than CC and MLO, and those with missing BI-RADS scores and a cancer status confirmed by biopsy. In addition, images with visible breast implants, images acquired with multiple manufacturers at the same screening visit, and those belonging to an incomplete or a non-screening exams were excluded. Appendix Figure~\ref{f:FlowChart} provides the data selection flowchart of this study.
\subsection{The segmentation model}

We employed two successive, task-specific convolutional neural networks based on modified UNet architectures, one for breast segmentation and the other for dense tissue segmentation. While both networks share the same architecture, they are trained independently and optimized for their respective tasks, resulting in different learned weights.

The original UNet architecture consists of two main components:\cite{Ronneberger_UNet_2015} a contracting path (encoder) and an expansive path (decoder), forming a symmetric structure with a total of 23 layers. The encoder captures contextual information through repeated applications of convolutional layers followed by max-pooling operations, progressively reducing spatial dimensions and increasing feature abstraction. The decoder then reconstructs spatial resolution using transposed convolutions, enabling precise localization. Crucially, skip connections are used to transfer feature maps from the encoder to the decoder at corresponding resolutions, allowing the network to leverage both high- and low-level features and thereby minimizing information loss during down-sampling.

To enhance training stability and efficiency, we replaced the standard UNet encoder with a ResNet-34 backbone.\cite{he_why_2019} This residual network architecture incorporates shortcut (or skip) connections, which help mitigate vanishing or exploding gradient issues by facilitating gradient flow during back-propagation. The ResNet-34 backbone consists of 34 convolutional layers organized into residual blocks, with the shortcut connections adjusted to match the spatial dimensions of the main computational path, ensuring consistent feature integration.

\subsection{Training dataset}

To fine-tune the segmentation model, a well-balanced dataset comprising 832 images from 208 randomly selected cancer-free women was used. The selection ensured balance across various factors, including centers, manufacturers, BI-RADS classes, and imaging views. Each woman contributed with a single complete screening exam, and ground-truth breast and dense masks were meticulously inspected by a medical reader and a methodologist, using a specialized tool for precise delineation of breast and dense areas developed by Therapixel, a French AI-focused company\cite{TPX_url}. In breast masks, the background and the pectoral muscle were assigned to class 0, whereas the breast represented class 1. Regarding the dense mask, class 0 corresponded to the non-dense tissue within the breast area, and class 1 defined the dense tissue.

The dataset was partitioned into training and test sets, maintaining a 9:1 ratio while preserving the earlier-mentioned balance across variables. This division was executed at the level of individual women, resulting in 748 images (187 women) allocated to the training set and 84 images (21 women) constituting the test set. The training set underwent an additional division into training and validation sets through a 10-fold cross-validation process detailed in the subsequent Section~\ref{sec::trainingstrategy}.

\subsection{Image preprocessing}

To ensure a standardized shape across all images, each original mammogram was resized along the height and cropped or padded along the width, achieving a targeted shape of 576x416. Next, pixel intensities were rescaled to [0,1] using the Values Of Interest Lookup Table (VOI LUT) transformation, and the background value was set to 0 for easier identification. These images constituted the inputs to the breast segmentation model, whereas the dense tissue segmentation model's inputs, containing only breast pixels, were obtained by applying the predicted breast mask to the preprocessed mammogram.

\subsection{Training strategy}
\label{sec::trainingstrategy}

We implemented our model using the lightweight PyTorch Lightning library for script simplicity and user-friendliness.\cite{falcon_pytorch_2019} Maintaining the original architecture of MammoFL, both UNets underwent training using identical hyperparameters, including a batch size of 16, a learning rate of $10^{-4}$, and no weight decay. The training was initiated with the pre-trained MammoFL version, achieving good performance after only ten epochs. The Adam optimizer was applied, and no data augmentation was introduced.

Model performance was evaluated using two complementary metrics:
\paragraph{Dice Similarity Coefficient (DSC)\cite{reinke_common_2023}}: An overlap-based metric that quantifies the agreement between predicted (pred) and reference (ref) segmentation masks
    \begin{equation} \label{eq:DSC}
    DSC\ (ref, \ pred) = \frac{2 \ |ref \ \cap \ pred|}{|ref|\ +\ |pred|}
    \end{equation}
    where $|ref|$ denotes the number of pixels in the reference mask, and $|ref \ \cap \ pred|$ represents the number of overlapping pixels between the prediction and the reference. DSC values range from 0 to 1, with higher values indicating better agreement. Training was performed by minimizing the Dice loss, defined as $loss = 1 - DSC$.
\paragraph{Normalized surface distance (NSD)\cite{reinke_common_2023}}: A boundary-based metric that measures the extent to which the predicted and reference boundaries overlap, accounting for a specified tolerance $\tau$:
    \begin{equation} \label{eq:NSD}
    NSD\ (ref, \ pred) = \frac{|S_{ref} \ \cap \ \mathcal{B}^{(\tau)}_{pred}| + |S_{pred} \ \cap \ \mathcal{B}^{(\tau)}_{ref}|}{|S_{ref}|\ +\ |S_{pred}|}
    \end{equation}
    where $S_{ref}$ and $S_{pred}$ are the boundaries of the reference and predicted masks, respectively, and $\mathcal{B}^{(\tau)}$ denotes the border region within $\tau$ pixels of the respective boundary. We set $\tau = 2$ pixels. NSD values range from 0 to 1, with values closer to 1 indicating better alignment of boundaries.

A joint training strategy was employed, prioritizing the achievement of the lowest validation dense loss, considering the higher complexity of dense area segmentation compared to breast segmentation. Moreover, ten-fold cross-validation was conducted to assess model performance. In each fold, data were randomly split into training and validation sets, respecting a 9:1 ratio. After each fold's training, weights yielding the lowest validation dense loss were saved. The simple unweighted average of these ten best weights was calculated to obtain the model version for evaluation on the holdout test set. This approach was preferred over selecting the weights yielding the lowest validation dense loss in a single fold to avoid favoring one fold and to ensure an averaged model behavior across \textit{all} training data. Appendix Figure~\ref{f:CrossVal} summarizes the training process.

\subsection{Inference step}
\label{sec::inferencestep}
Using our deep learning model, we calculated dense area and percent density across a dataset of 1,262,276 images corresponding to 315,569 visits from 77,298 women. Selection criteria for the inference dataset included women with at least two screening visits, aged between 40 and 74 years at their initial visit. Refer to Appendix Figure~\ref{f:FlowChart} for more details.

Our approach to synthesizing quantitative density metrics at the screening visit level involved a two-step process. First, acknowledging that CC and MLO views provide specific information, we computed the average value between these views for the same breast. Then, for dense area, we calculated the sum of the dense areas in both breasts to depict the total amount of dense tissue within the individual. For percent density, the total assessment at each visit was the average value between the two breasts, ensuring balanced representation.
\section{The joint model for density association with breast cancer risk}
\label{sec::JM_method}

To analyze the longitudinal evolution of quantitative density metrics and the risk of breast cancer, accounting for their mutual correlation, informative dropouts, and irregular intervals between visits, a shared-parameter joint model is employed.

\subsection{The longitudinal sub-model}
\label{sec::JM_longi}

Let $N$ be the number of women participating in a breast cancer screening program. In the US, annual screening is recommended for women at average risk for breast cancer starting at age 40. However, from the age of 45 onward, regular annual screening is strongly advised.\cite{oeffinger_breast_2015}

Quantitative mammographic density metrics represent the longitudinal biomarker (dense area and percent density), where $\boldsymbol{Y_{i}} = (Y_{i}(t_{i1})^\top, Y_{i}(t_{i2})^\top,\ldots, Y_{i}(t_{in_{i}})^\top)^\top$ denotes the vector of measurements for woman $i$ ($i = 1,\ldots,N$) at times $t_{ij}$ ($j=1,\ldots,n_i$). Breast cancer diagnosis interrupts the screening period and thus the collection of the biomarker's measurements.
The observed value of the biomarker is assumed to be noisy, and thus, its true value $m_{ij}$, remains unobserved. A linear mixed-effects model is used to describe the trajectory over time of the biomarker $Y_{i}(t_{ij}) = Y_{ij}$ as follows
\begin{equation} \label{eq:Y_ij}
    \begin{array}{ll}
        Y_{ij} & = m_{ij} + \epsilon_{ij} \\
                &  = \boldsymbol{X_{ij}}^{D^\top} \ \boldsymbol{\beta} + \boldsymbol{Z_{ij}}^{D^\top} \ \boldsymbol{b_{i}} + \epsilon_{ij}
    \end{array}
\end{equation}
with $\boldsymbol{X_{ij}^{D}}$ and $\boldsymbol{Z_{ij}^{D}}$ two vectors of covariates associated to $\boldsymbol{\beta}$ the p-vector of fixed effect parameters, and $\boldsymbol{b_i}$ the q-vector ($ q \leq p$) of normally distributed individual random effects ($\boldsymbol{b_i} \sim \mathcal{N}(0, \boldsymbol{B})$), respectively. The superscript $D$ indicates that these covariates and effects pertain to the longitudinal submodel associated with mammographic density. Measurement errors $\epsilon_{ij}$ are independent, normally distributed ($\epsilon_{ij} \sim \mathcal{N}(0, \sigma_{\epsilon}^{2})$), and independent from $\boldsymbol{b_i}$.

\subsection{The survival sub-model}
\label{sec::JM_surv}
Let $T^{*}_{i}$ be the time-to-breast cancer for a woman $i$ ($i=1,...,N$), with breast cancer event confirmed through a positive biopsy exam following a suspicious screening visit. In our survival model, age is used as the time scale to account for delayed entry and left truncation.\cite{rondeau_frailtypack_2012} Indeed, women with an average risk of breast cancer may start their screening at 40 or later, and those with a prior occurrence of the event before screening initiation are excluded. Therefore, the time-to-breast cancer is defined as the time elapsed between the woman’s age at first screening mammography ($t_{0i}$) and the age at diagnosis confirmed by biopsy. Given that some women’s follow-up may be censored before the occurrence of breast cancer, the observed time-to-event is defined as $T_{i} = min(T^{*}_{i}, C_{i})$, where $C_{i}$ is the non-informative right censoring time, and $\delta_i$ is the event indicator with $\delta_{i}=1$ when breast cancer is diagnosed before censoring ($T^{*}_{i} \leq C_{i}$), and $\delta_{i}=0$ otherwise.

The time-to-breast cancer $T_{i}$ is modeled using a proportional hazards model. It is assumed that the hazard of the event occurrence depends on an individual-specific profile of mammographic density. In the presence of delayed entry and left truncation, the individual hazard function equals zero before $t_{0i}$, and for all $t > t_{0i}$ is defined as
\begin{equation} \label{eq:survival_submod}
    h_{i}(t) = h_{0}(t) \exp\Big{(} \boldsymbol{X_{i}}^{C^\top} \ \boldsymbol{\gamma} + \ f(\cdot)^\top \ \boldsymbol{\alpha}  \Big{)},\ \forall \ t > t_{0i}
\end{equation}
where $h_{0}(t)$ denotes the unspecified baseline hazard function common to all subjects, $\boldsymbol{\gamma}$ is the coefficient vector associated to the vector $\boldsymbol{X_{i}^{C}}$ corresponding to baseline risk factors for breast cancer, and the function $f()$ and its vector of parameters $\boldsymbol{\alpha}$ that quantifies the association between the longitudinal biomarker information and the time-to-event outcome. Four functions $f(\cdot)$ are specified as follows:
\paragraph{The current level}: The instantaneous risk of breast cancer is directly associated with the current 'true' level of mammographic density at time $t$.
    \begin{equation} \label{eq:CV}
    f(\cdot) = \alpha_{1} \ m_{i}(t)
    \end{equation}
    A positive $\alpha_{1}$ indicates that a 1-unit increase in the longitudinal biomarker at time $t$ corresponds to an increase in the log hazard ratio of breast cancer at the same time $t$.
\paragraph{The current slope}: The instantaneous risk of the event occurrence is associated with the rate of change of the current 'true' individual trajectory (or slope) of mammographic density at time $t$.
    \begin{equation} \label{eq:CS}
    f(\cdot) = \alpha_{2} \ m_{i}^{'}(t)
    \end{equation}
    A positive and significant $\alpha_{2}$ indicates that the steeper the current slope at time $t$ (i.e., the biomarker is increasing rapidly), the higher the risk of the event at the same time $t$.
\paragraph{The combination of current level and slope}: The instantaneous risk of breast cancer is associated to both the current level and slope of mammographic density at time $t$.
    \begin{equation} \label{eq:CVCS}
    f(\cdot) = \alpha_{1} \ m_{i}(t) + \alpha_{2} \ m_{i}^{'}(t)
    \end{equation}
\paragraph{The cumulative effect}: The effect of the cumulative mammographic density level on the breast cancer hazard over time is quantified by $\alpha_{3}$ by integrating its longitudinal trajectory from an individual initial time $t_{0i}$ up to the current time $t$ divided by $t$ to account for the observation period.
    \begin{equation} \label{eq:Area}
    f(\cdot) = \alpha_{3} \ \frac{\int_{t_{0i}}^{t} m_{i}(s) ds}{t}
    \end{equation}
    Here, a positive $\alpha_{3}$ signifies that a 1-unit increase in the area under the longitudinal trajectory of mammographic density results in an increased log hazard ratio of the event, considering the entire longitudinal profile of the biomarker rather than just its current level or slope at time $t$.

\subsection{Bayesian inference}
Let $\boldsymbol{\theta}=(\boldsymbol{\beta}, \boldsymbol{\gamma}, \boldsymbol{\alpha}, \sigma_{\epsilon}^{2}, \boldsymbol{B})$ be the vector of the joint model's parameters, in addition to $h_{0}(t)$, the baseline hazard function, which is approximated using a penalised B-spline. Bayesian inference was used to estimate the joint model's parameters with the posterior distribution of $\boldsymbol{\theta}$. Given the prior distribution $p(\boldsymbol{\theta})$, the likelihood function $p(\textbf{E}|\boldsymbol{\theta}) = L(\boldsymbol{\theta})$, and the data (or evidence) $\textbf{E} = (\textbf{Y},\textbf{T},\boldsymbol{\delta})$, the posterior probability distribution of $\boldsymbol{\theta}$, $p(\boldsymbol{\theta}|\textbf{E})$, is defined as follows
\begin{equation} \label{eq:Bayesian_inf}
p(\boldsymbol{\theta}| \textbf{D}) = \frac{p(\textbf{D}|\boldsymbol{\theta})p(\boldsymbol{\theta})}{p(\textbf{D})} \propto p(\textbf{D}|\boldsymbol{\theta})p(\boldsymbol{\theta})
\end{equation}
where $p(\textbf{E})$ is the marginal likelihood. Markov chain Monte Carlo (MCMC) sampling methods, such as the Metropolis–Hastings algorithm, is used to estimate the posterior distribution.

Given the substantial scale of our dataset (Appendix Figure~\ref{f:FlowChart}), we experienced challenges related to prolonged computation time and limited memory resources when fitting the joint model. These challenges were eased using the consensus Monte Carlo algorithm\cite{afonso_efficiently_2023}, following a structured approach: First, we separated our dataset into 12 disjoint splits. Then, parallel independent MCMC simulations were run on each split, generating $c$ draws per chain for each parameter in $\boldsymbol{\theta}$. These MCMC samples were next combined across all splits using the precision weight method that assigns a split-specific weight to each MCMC draw in each chain. This weight reflected the precision in each posterior sub-sample, with greater precision resulting in a higher weight. The joint model was run using \texttt{JMbayes2} (v0.4-5)\cite{rizopoulos_jmbayes2_2023} by considering the package’s default prior distributions. We employed 3 Markov chains with $8,500$ iterations per chain, discarding the first $3,500$ iterations as a warm-up.
Chains convergence was assessed using the convergence diagnostic $\hat{R}$.
To select the optimal joint model that fits the data, we consider the Deviance Information Criterion (DIC)\cite{spiegelhalter_deviance_2014}, where lower DIC values indicate a superior fit of the model to the data.

\subsection{Individual risk predictions}
\label{sec::pred_method}
The joint model described above was used to calculate individual dynamic predictions. In particular, we aim to represent $\pi(s+w)$ the probability of experiencing the event of interest in a specific time span $[s, s+w]$ defined by a fixed window of prediction $w$ and a varying landmark time $s$ representing the time at which predictions are made conditionally to the subject-specific history. Given that the woman did not experience breast cancer before time $s$, where $s > t_{0i}$, the prediction is conditioned on the woman’s history $\boldsymbol{H_{i}(s)} = \{\boldsymbol{X_{i}^{C}}; \ Y_{i,1}, ..., Y_{i,k}|\ t_{0i} < k \leq s \}$ formed by $\boldsymbol{X_{i}^{C}}$ the baseline covariates and $\boldsymbol{Y_{i}}$ all the observed values of the biomarker before time $s$.\cite{krol_joint_2016} For all $s > t_{0i}$ under the Bayesian framework, the individual probability $\pi_{i}(s+w)$ is defined as
\begin{align} \label{eq:dynamicpred_surv}
    \pi_{i}(s+w)
    &= \mathcal{P}(T_{i} \leq s+w \mid T_{i} > s,
       \boldsymbol{H_{i}(s)},\boldsymbol{E}) \notag \\[0.2cm]
    &= \mathlarger{\iint}
       \frac{S_{i}(s \mid \boldsymbol{H_{i}(s)}, \boldsymbol{b_{i}}, \boldsymbol{\theta})
       - S_{i}(s+w \mid \boldsymbol{H_{i}(s)}, \boldsymbol{b_{i}}, \boldsymbol{\theta})}
       {S_{i}(s \mid \boldsymbol{H_{i}(s)}, \boldsymbol{b_{i}}, \boldsymbol{\theta})} \notag \\
    &\quad \times p(\boldsymbol{b_{i}} \mid T_{i} > s,\boldsymbol{H_{i}(s)}, \boldsymbol{\theta})
       \; p(\boldsymbol{\theta}\mid\textbf{E})
       \, d\boldsymbol{b_{i}} \, d\boldsymbol{\theta}
\end{align}
To compute $\pi_{i}(s+w)$, we employ a Monte Carlo scheme to approximate its posterior distribution. Initially, we sample values of $\tilde{\boldsymbol{\theta}}$ and $\tilde{\boldsymbol{b_{i}}}$ from the posterior distributions of the parameters $p(\boldsymbol{\theta}|\textbf{D})$ and the random effects $p(\boldsymbol{b_{i}}|T_{i} > s,\boldsymbol{H_{i}(s)}, \boldsymbol{\theta})$, respectively. Then, we calculate the ratio of survival probabilities as defined in Equation~\ref{eq:dynamicpred_surv}. By repeating this process $\textit{L}$ times, the prediction is estimated as the mean of all $\textit{L}$ probabilities. The $95\%$ credible interval is derived using the $2.5\%$ and $97.5\%$ percentiles of the posterior distribution's\cite{rizopoulos_jmbayes2_2023}. We assessed predictive accuracy using standard measures for right-censored data, including the area under the receiver operating characteristics curve (AUC), and the Brier score (BS) as defined in Blanche et al. (2015)\cite{blanche_quantifying_2015}. Additional details are provided in the Appendix~B. Internal validation was conducted using a patient-level k-fold cross-validation ($k = 5$) to correct for over-optimism. For each left-out fold, the individual predictions are derived from estimates obtained using the joint model built on the remaining $k-1$ folds\cite{krol_joint_2016}.
\section{Results}
\label{sec::results}

\subsection{Quantitative mammographic density estimation}

Table~\ref{t:MammoFL_DSCs} displays the evaluation metrics derived from the left-out test set, which included 84 images from 21 women. The model performed well overall, achieving an average DSC and NSD of 0.993 (95\% confidence interval (95\%CI): 0.991, 0.994) and 0.960 (95\%CI:0.945, 0.972) for breast and 0.848 (95\%CI: 0.822, 0.873) and 0.728 (95\%CI: 0.709, 0.747) for dense areas, respectively. The performance was consistent across different views, manufacturers, and screening centers. While breast DSC values remained high across all BI-RADS categories, dense DSC showed some variability, particularly in the A category, though it stayed within acceptable limits. Figure~\ref{f:example_dl_perf} illustrates the segmentation results in processed mammograms across different BI-RADS categories and manufacturers for four subjects, highlighting the model's ability to accurately delineate breast and dense areas with good DSC values.

\begin{table}[htbp]
\caption{\textbf{Performance of the deep learning model in segmenting breast and dense areas on the test set (\boldsymbol{$N=84$} images).} Data represent mean (95\% confidence intervals) of Dice Similarity Coefficients and Normalized Surface Distances of breast and dense UNets.}
\label{t:MammoFL_DSCs}
\begin{center}
\resizebox{\textwidth}{!}{%
\begin{tabular}{l|cc|cc}
\hline
 & \multicolumn{2}{|c|}{\textbf{Breast UNet}} & \multicolumn{2}{c}{\textbf{Dense UNet}} \\
 & \textbf{DSC} & \textbf{{NSD}} & \textbf{DSC} & \textbf{{NSD}} \\
\hline
\textbf{Image-level ($N_{image}$ = 84)} & \textbf{0.993 (0.991, 0.994)} & \textbf{0.960 (0.949, 0.970)} & \textbf{0.848 (0.822, 0.873)} & \textbf{0.728 (0.709, 0.747)}\\
\hline
\textbf{Center} & & & &\\
 \quad 1 ($N_{image}$ = 64) & 0.993 (0.991, 0.994) & 0.960 (0.945, 0.972) & 0.845 (0.817, 0.873) & 0.721 (0.701, 0.740) \\
 \quad 2 ($N_{image}$ = 20) & 0.993 (0.990, 0.995) & 0.960 (0.941, 0.974) & 0.859 (0.808, 0.902) & 0.751 (0.700, 0.798) \\
\textbf{Manufacturer} & & & & \\
 \quad Hologic ($N_{image}$ = 52) & 0.993 (0.991, 0.994) & 0.964 (0.953, 0.974) & 0.849 (0.817, 0.881) & 0.730 (0.702, 0.759) \\
 \quad GEHC ($N_{image}$ = 32) & 0.993 (0.989, 0.995) & 0.953 (0.931, 0.972) & 0.847 (0.808, 0.881) & 0.725 (0.700, 0.748) \\
 \textbf{View} & & & & \\
 \quad CC ($N_{image}$ = 42) & 0.996 (0.995, 0.997) & 0.983 (0.975, 0.989) & 0.840 (0.801, 0.876) & 0.737 (0.707, 0.766) \\
 \quad MLO ($N_{image}$ = 42) & 0.990 (0.987, 0.992) & 0.937 (0.919, 0.952) & 0.857 (0.825, 0.887) & 0.720 (0.693, 0.745)\\
\textbf{BI-RADS score} & & & & \\
 \quad A ($N_{image}$ = 20) & 0.995 (0.992, 0.997) & 0.966 (0.992, 0.997) & 0.675 (0.636, 0.714) & 0.628 (0.593, 0.664) \\
 \quad B ($N_{image}$ = 20) & 0.991 (0.988, 0.995) & 0.955 (0.988, 0.994) & 0.841 (0.817, 0.860) & 0.735 (0.716, 0.752) \\
 \quad C ($N_{image}$ = 24) & 0.994 (0.992, 0.995) & 0.969 (0.992, 0.995) & 0.916 (0.905, 0.928) & 0.763 (0.738, 0.787) \\
 \quad D ($N_{image}$ = 20) & 0.990 (0.986, 0.994) & 0.946 (0.985, 0.994) & 0.947 (0.937, 0.957) & 0.780 (0.743, 0.817) \\
\hline
\textbf{Patient-level (N = 21)} & \textbf{0.993 (0.991, 0.995)} & \textbf{0.960 (0.951, 0.970)} & \textbf{0.848 (0.814, 0.885)} & \textbf{0.728 (0.704, 0.751)}\\
\hline
\multicolumn{5}{l}{DSC: Dice Similarity Coefficient; NSD: Normalized Surface Distance; GEHC: General Electric Healthcare}\\
\multicolumn{5}{l}{CC: craniocaudal; MLO: mediolateral oblique; BI-RADS: breast imaging reporting and data system}\\
\end{tabular}%
}
\end{center}
\end{table}

After completing the training phase of the deep learning model, we proceeded to the inference phase, including a total of 77,298 women. Table~\ref{t:Women_chars} outlines their main characteristics. The median follow-up duration was 4.0 [IQR: Interquartile range 2.1--6.3] years, with median mammography visits occurring every 12.2 [IQR: 11.4--14.8] months. Most women (68.7\%) underwent three or more mammography rounds, with the maximum number reaching 13. Over the screening period from 2006 to 2019, 979 (1.3\%) women had breast cancer.

\begin{table}
\caption{\textbf{Baseline characteristics of women included in the analysis of the association between longitudinal mammographic density and breast cancer risk using a joint model.}}
\label{t:Women_chars}
\begin{center}
\begin{small}
\begin{tabular}{lccc}
\hline
\textbf{Variable} & \multicolumn{2}{c}{\textbf{Breast cancer status}} & \textbf{Total} \\
 & \textbf{Cancer-free} & \textbf{Cancer} & \\
  & \textbf{N = 76,319} & \textbf{N = 979} & \textbf{N = 77,298} \\
\hline
\textbf{Age at first visit, y} & & & \\
& 52 [45--61]  & 57 [49--64] & 52 [45--61] \\
\textbf{Number of visits} & & & \\
 \quad Two & 23,892 (31.3) & 288 (29.4) & 24,180 (31.3) \\
 \quad Three or more & 52,427 (68.7) & 691 (70.6) & 53,118 (68.7) \\
\textbf{Center} & & & \\
 \quad 1 & 49,244 (64.5) & 670 (68.4) & 49,914 (64.6)  \\
 \quad 2 & 27,075 (35.5) & 309 (37.6) & 27,384 (35.4) \\
\textbf{BI-RADS score} & & & \\
 \quad A & 8,958 (11.8) & 76 (7.8) & 9,034 (11.7) \\
 \quad B & 27,645 (36.2) & 343 (35.0) & 27,988 (36.2) \\
 \quad C & 32,288 (42.3) & 465 (47.5) & 32,753 (42.4) \\
 \quad D & 7,428 (9.7) & 95 (9.7) & 7,523 (9.7) \\
 \textbf{Dense area, $cm^{2}$} & & & \\
& 78.1 [44.4--115.0] & 92.2 [55.8--127.9] & 77.9 [44.3--114.8] \\
 \textbf{Percent density, $\%$} & & & \\
& 34.5 [16.0--50.0] & 37.0 [20.1--49.6] & 34.5 [16.0--50.0] \\
\hline
\multicolumn{4}{l}{Data are median [IQR] and n (\%)}\\
\multicolumn{4}{l}{BI-RADS: breast imaging reporting and data system}\\
\end{tabular}
\end{small}
\end{center}
\end{table}

\begin{figure}[htbp]
 \centerline{\includegraphics[width=6.8in]{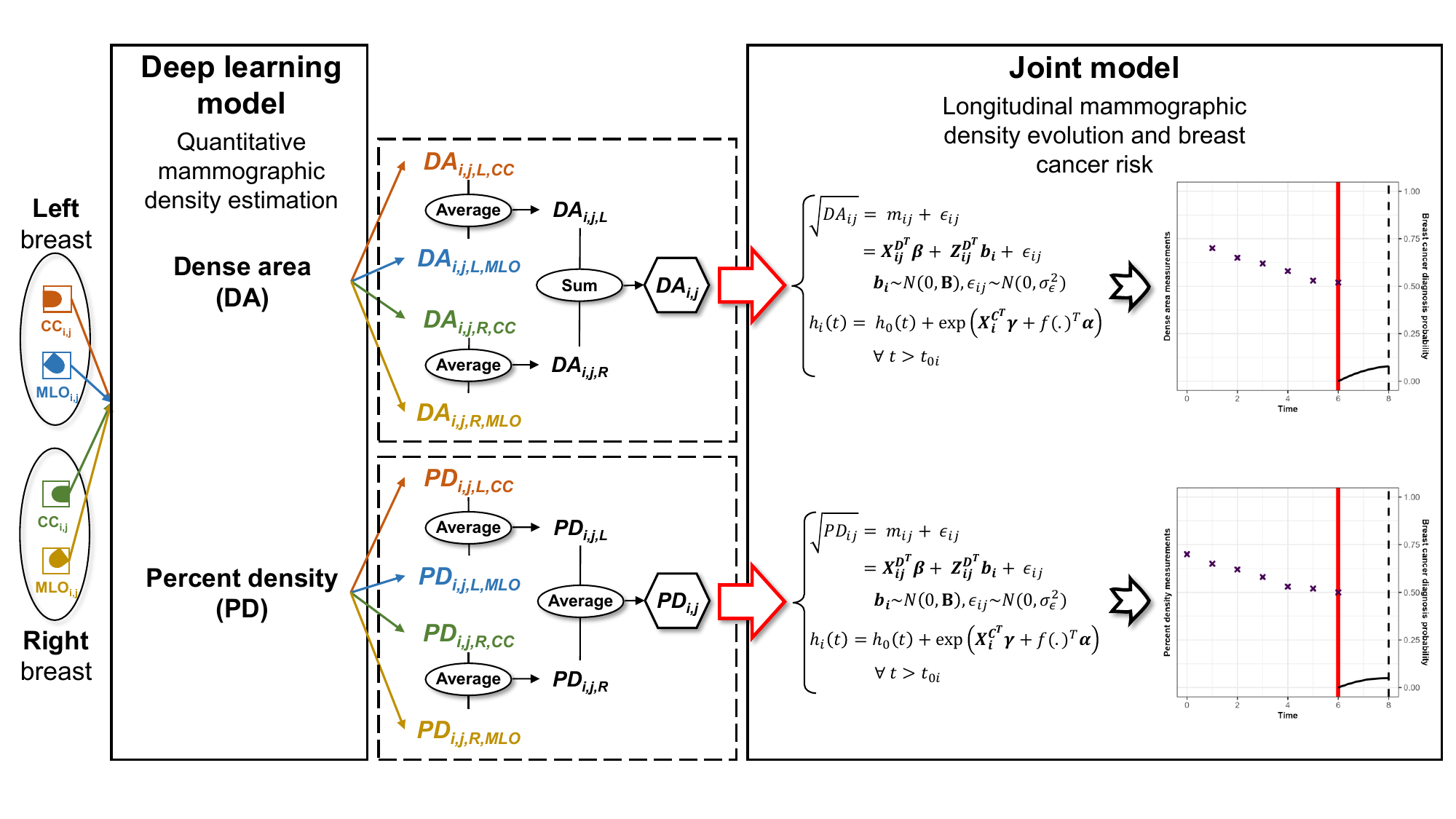}}
\caption{\textbf{The DeepJoint algorithm pipeline.} For each woman $i$ at screening visit $t_{ij}$, the complete exam (four images: LCC, LMLO, RCC, RMLO) is considered. A deep learning segmentation model estimates dense area (DA) and percent density (PD) for each view and each breast. These measures are then aggregated: DA is summed across breasts to represent total dense tissue, while PD is averaged across views and breasts. The resulting visit-level quantitative biomarkers are used as longitudinal inputs in a shared-parameter joint model that jointly models density trajectories and time-to-breast cancer, capturing their association over time and enabling individual and dynamic risk predictions.}
\label{f:longi_biomarker}
\end{figure}

We derived visit-level density metrics as described in Section~\ref{sec::inferencestep} and Figure~\ref{f:longi_biomarker} provides a visual representation of this process and its integration into the DeepJoint algorithm.

Baseline dense area estimates ranged from 0 to 459.05 $cm^{2}$, with median values of 77.9 [IQR: 44.3--114.8] $cm^{2}$ in cancer-free women and 92.2 [IQR: 55.8--127.9] $cm^{2}$ in those diagnosed with breast cancer. Baseline percent density estimates varied from 0 to 93.3\% with medians of 34.5 [IQR: 16.0--50.0] \% and 37.0 [IQR: 20.1--49.6] \% in cancer-free and cancer groups, respectively. Appendix Table~A.3 reports baseline dense area and percent density estimates according to BI-RADS scores. Our deep learning model provided adequate quantitative mammographic density estimations across BI-RADS categories, with visually distinguishable patterns that correspond closely to the four BI-RADS density groups. Women diagnosed with breast cancer generally had higher median values compared to those without breast cancer, except for median percent density values in C and D classes. Although there was a slight difference, percent density assessments in these classes remained very close.

\subsection{Longitudinal mammographic density association with breast cancer}
To ensure normality, we considered the square root of our longitudinal biomarkers in the longitudinal sub-model. Appendix Figure~\ref{f:Spaghetti} displays the square root of the longitudinal biomarkers measurements for 1,000 randomly selected women, connected by lines. Using a natural cubic spline, data points were smoothed and represented by the red curve with 95\% confidence bands illustrated in black. Overall, both biomarkers seem to decrease with age, as described in the literature\cite{burton_mammographic_2017}.

We fitted two distinct joint models, each using one of the two longitudinal biomarkers ($Y_{i}$ is either dense area or percent density). This strategy allowed us to (i) explore, for the first time, the longitudinal association between percent density and breast cancer risk, (ii) evaluate the clinical relevance of each newly derived biomarker from our deep learning model, and (iii) disentangle their individual contributions to risk and show distinct effect magnitudes, if any. We addressed delayed entry and left truncation by adjusting for age, as explained in Section~\ref{sec::JM_surv}. Additionally, we included age at the first screening visit ($age_{0}$) and manufacturer ($manuf$) as additional factors in the longitudinal sub-model. Women with missing covariates or cancer status were excluded. We evaluated four link functions, as reported in Equations ~\ref{eq:CV}, ~\ref{eq:CS}, ~\ref{eq:CVCS} and~\ref{eq:Area}.
\begin{equation} \label{eq:JM_general}
    \left\{
        \begin{array}{ll}
            \sqrt{Y_{ij}} & = m_{ij}  + \epsilon_{ij}\\[0.1cm]
            & = (\beta_{0} + b_{i0}) + (\beta_{1} +   b_{i1}) \ age_{ij} + \beta_{2} \ age_{0i} + \ \beta_{3} \ manuf_{i} + \epsilon_{ij} \\[0.2cm]
            h_{i}(age) & = h_{0}(age) \ exp\Big{\{}\alpha \ f(m_{i}(age))\Big{\}} \\[0.2cm]
        \end{array}
    \right.
\end{equation}
Results for all fitted joint models are presented in Table~\ref{t:JM_DA_PD}. These models exhibited satisfactory convergence properties with $\hat{R} < 1.10$ for all coefficients. Regardless of the biomarker, the joint model with the current level and slope of the biomarker's trajectory at a given age had the best fit to data according to the DIC values (Model 3: dense area, DIC = 82,070.83; percent density, DIC = 58,304.56).

Results were consistent across all joint models, indicating a negative association between the longitudinal biomarker and age, aligning with existing literature\cite{burton_mammographic_2017} and the observed trends in Appendix Figure~\ref{f:Spaghetti}. The association between the longitudinal biomarker and the age at the first screening visit ($age_{0}$) varied depending on the biomarker: a positive association was observed for dense area across all models, while no association was observed for percent density. Additionally, a positive association was identified between the longitudinal biomarker and Hologic manufacturers, regardless of the biomarker type.

Furthermore, a positive association was observed between the longitudinal dense area's current level and slope and the risk of breast cancer, with coefficients of 0.116 ($95\%$ credible interval (CI): 0.056, 0.174) and 0.042 (0.019, 0.062), respectively. Similar patterns were observed in the joint model incorporating percent density as the longitudinal biomarker, where the current level and slope of the biomarker's trajectory were positively associated with the risk of breast cancer, yielding coefficients of 0.134 (0.041, 0.228) and 0.038 (0.018, 0.053), respectively. These findings suggest that elevated levels of quantitative mammographic density at a given age are associated to a higher risk of breast cancer. Moreover, in the context of a continuous decrease in mammographic density over time, women with a slower decrease (i.e. a negative slope close to zero) in their biomaker's level are more prone to breast cancer risk compared to those with a faster decrease (i.e., a steeper negative slope), assuming similar levels of the biomarker at a specific age.

Consistent trends were also noted in Models 1 and 2 (Table~\ref{t:JM_DA_PD}) where either the current value or the current slope of the longitudinal biomarker were incorporated, mirroring the outcomes of the best model. Additionally, Model 4, employing the cumulative mammographic density level up to a definite age as a link function, also suggested a positive association between the cumulative mammographic density level and the risk of breast cancer.
\begin{table}
\caption{\textbf{Estimated posterior means and 95\% credible intervals (CIs) for joint model coefficients in a dataset of $N = 77,298$ women.} The models utilize different link functions, including the current level (Model 1), the current slope (Model 2), the current level and slope (Model 3), and the cumulative level (Model 4) of the longitudinal biomarker to assess their association with breast cancer risk.}
\label{t:JM_DA_PD}
\begin{center}
\resizebox{\textwidth}{!}{%
\begin{tabular}{lcccccc}
\hline
& \multicolumn{3}{c}{\textbf{JM with} $\boldsymbol{\sqrt{dense\ area\ (cm^{2})}}$} & \multicolumn{3}{c}{\textbf{JM with} $\boldsymbol{\sqrt{percent\ density\ (\%)}}$}\\
\textbf{Coefficient} & Mean & 95$\%$CI & $\hat{R}$ & Mean & 95$\%$CI & $\hat{R}$ \\
\hline
\textbf{Model 1} & & & & & & \\
\quad \textbf{\textit{Longitudinal sub-model}} & & & & & & \\
 \qquad Intercept ($\beta_{0}$) & 9.135 & (8.839, 9.432) & 1.006 & 6.408 & (6.207, 6.612) & 1.002 \\
 \qquad Age ($\beta_{1}$) & -0.114 & (-0.118, -0.111) & 1.029 & -0.081 & (-0.084, -0.079) & 1.002 \\
 \qquad $Age_{0}$ ($\beta_{2}$) & 0.017 & (0.012, 0.023) & 1.013 & 0.000 & (-0.004, 0.004) & 1.006  \\
 \qquad $Manuf.^{*}$ ($\beta_{3}$) & 0.163 & (0.145, 0.182) & 1.022 & 0.184 & (0.172, 0.196) & 1.035 \\
 \qquad $\sigma_{\epsilon}$ & 0.603 & (0.599, 0.606) & 1.002 & 0.375 & (0.373, 0.377) & 1.003 \\
\quad \textbf{\textit{Survival Sub-model}} & & & & & & \\
 \qquad Current level ($\alpha_{1}$)& 0.109 & (0.051, 0.166) & 1.000 & 0.125 & (0.033, 0.219) & 1.003 \\
\quad \textbf{\textit{DIC}} & \multicolumn{3}{c}{82,092.46} & \multicolumn{3}{c}{58,329.67}\\
\hline
\textbf{Model 2} & & & & & & \\
\quad \textbf{\textit{Longitudinal sub-model}} & & & & & & \\
 \qquad Intercept ($\beta_{0}$)  & 9.144 & (8.846, 9.440) & 1.004 & 6.412 & (6.214, 6.611) & 1.001 \\
 \qquad Age ($\beta_{1}$) & -0.114 & (-0.117, -0.111) & 1.025 & -0.081 & (-0.084, -0.079) & 1.021 \\
 \qquad $Age_{0}$ ($\beta_{2}$) & 0.017 & (0.011, 0.023) & 1.010 & 0.000 & (-0.004, 0.004) & 1.004  \\
 \qquad $Manuf.^{*}$ ($\beta_{3}$) & 0.161 & (0.142, 0.181) & 1.041 & 0.184 & (0.172, 0.196) & 1.035 \\
 \qquad $\sigma_{\epsilon}$ & 0.603 & (0.599, 0.607) & 1.003 & 0.375 & (0.373, 0.377) & 1.006 \\
\quad \textbf{\textit{Survival Sub-model}} & & & & & & \\
 \qquad Current slope ($\alpha_{2}$) & 0.038 & (0.004, 0.063) & 1.003 & 0.033 & (0.006, 0.053) & 1.013 \\
\quad \textbf{\textit{DIC}}  & \multicolumn{3}{c}{82,097.67} & \multicolumn{3}{c}{58,326.79}\\
\hline
\textbf{Model 3} & & & & & & \\
\quad \textbf{\textit{Longitudinal sub-model}} & & & & & & \\
 \qquad Intercept ($\beta_{0}$) & 9.155 & (8.857, 9.453) & 1.009 & 6.410 & (6.213, 6.608) & 1.007 \\
 \qquad Age ($\beta_{1}$) & -0.114 & (-0.118, -0.110) & 1.022 & -0.082 & (-0.084, -0.079) & 1.019 \\
 \qquad $Age_{0}$ ($\beta_{2}$) & 0.017 & (0.011, 0.023) & 1.013 & 0.000 & (-0.003, 0.004) & 1.011  \\
 \qquad $Manuf.^{*}$ ($\beta_{3}$) & 0.162 & (0.143, 0.181) & 1.009 & 0.184 & (0.172, 0.195) & 1.005 \\
 \qquad $\sigma_{\epsilon}$ & 0.603 & (0.599, 0.607) & 1.056 & 0.375 & (0.373, 0.378) & 1.076 \\
\quad \textbf{\textit{Survival Sub-model}} & & & & & & \\
 \qquad Current level ($\alpha_{1}$) & 0.116 & (0.056, 0.174) & 1.002 & 0.134 & (0.041, 0.228) & 1.005 \\
 \qquad Current slope ($\alpha_{2}$) & 0.042 & (0.019, 0.062) & 1.003 & 0.038 & (0.018, 0.053) & 1.004\\
\quad \textbf{\textit{DIC}}  & \multicolumn{3}{c}{\textbf{82,070.83}} & \multicolumn{3}{c}{\textbf{58,304.56}}\\
\hline
\textbf{Model 4} & & & & & & \\
\quad \textbf{\textit{Longitudinal sub-model}} & & & & & & \\
 \qquad Intercept ($\beta_{0}$)& 9.147 & (8.849, 9.447) & 1.000 & 6.414 & (6.214, 6.611) & 1.002 \\
 \qquad Age ($\beta_{1}$) & -0.114 & (-0.118, -0.111) & 1.002 & -0.081 & (-0.084, -0.079) & 1.002 \\
 \qquad $Age_{0}$ ($\beta_{2}$) & 0.017 & (0.011, 0.023) & 1.001 & 0.000 & (-0.003, 0.004) & 1.004  \\
 \qquad $Manuf.^{*}$ ($\beta_{3}$) & 0.163 & (0.144, 0.182) & 1.012 & 0.184 & (0.172, 0.196) & 1.011 \\
 \qquad $\sigma_{\epsilon}$ & 0.603 & (0.599, 0.606) & 1.016 & 0.375 & (0.372, 0.377) & 1.003 \\
\quad \textbf{\textit{Survival Sub-model}} & & & & & & \\
 \qquad Cumulative level ($\alpha_{3}$)& 0.125 & (0.064, 0.184) & 1.001 & 0.149 & (0.054, 0.245) & 1.002 \\
\quad \textbf{\textit{DIC}}  & \multicolumn{3}{c}{82,093.73} & \multicolumn{3}{c}{58,328.89}\\
\hline
\multicolumn{5}{l}{$Manuf.^{*}$: Hologic vs. GEHC manufacturer; JM: Joint model; CI: Credible interval}\\
\multicolumn{5}{l}{DIC: Deviance information criterion; Bold DICs represent the best joint model}\\
\multicolumn{5}{l}{$\hat{R}$ for chains convergence diagnostic ($\hat{R} \leq 1.1$ means good convergence) }\\
\end{tabular}%
}
\end{center}
\end{table}

\subsection{Breast cancer prediction}
Individual and dynamic risk predictions in the next five years were computed using the optimal joint model (Model 3, Table~\ref{t:JM_DA_PD}). We illustrate the results for 24 randomly selected women, grouped according to age at screening initiation: 42 years (Group 1), 54 years (Group 2), and 65 years (Group 3). For each group, we evaluate 5-year breast cancer risk at landmark ages within the following intervals: 45–49 years for Group 1, 58–62 years for Group 2, and 68–72 years for Group 3. Results for Group 2 are presented in Figure~\ref{f:dyn_pred_age54_w5}, and other group's figures are provided in the Appendix. Comparing age groups, breast cancer risk seemed to increase with age for women with similar dense area or percent density profiles. Within an age group, women with consistently high mammographic density levels over time exhibited higher 5-year breast cancer risks, coherent with results in Table~\ref{t:JM_DA_PD}. Also, individual 5-year risk probabilities differed when using either dense area or percent density as the longitudinal biomarker for some women. We also computed individual 2-year breast cancer risk predictions for all groups to align with screening visit frequencies, yielding similar results (Appendix).

\begin{figure}[htbp]
 \centerline{\includegraphics[width=7in]{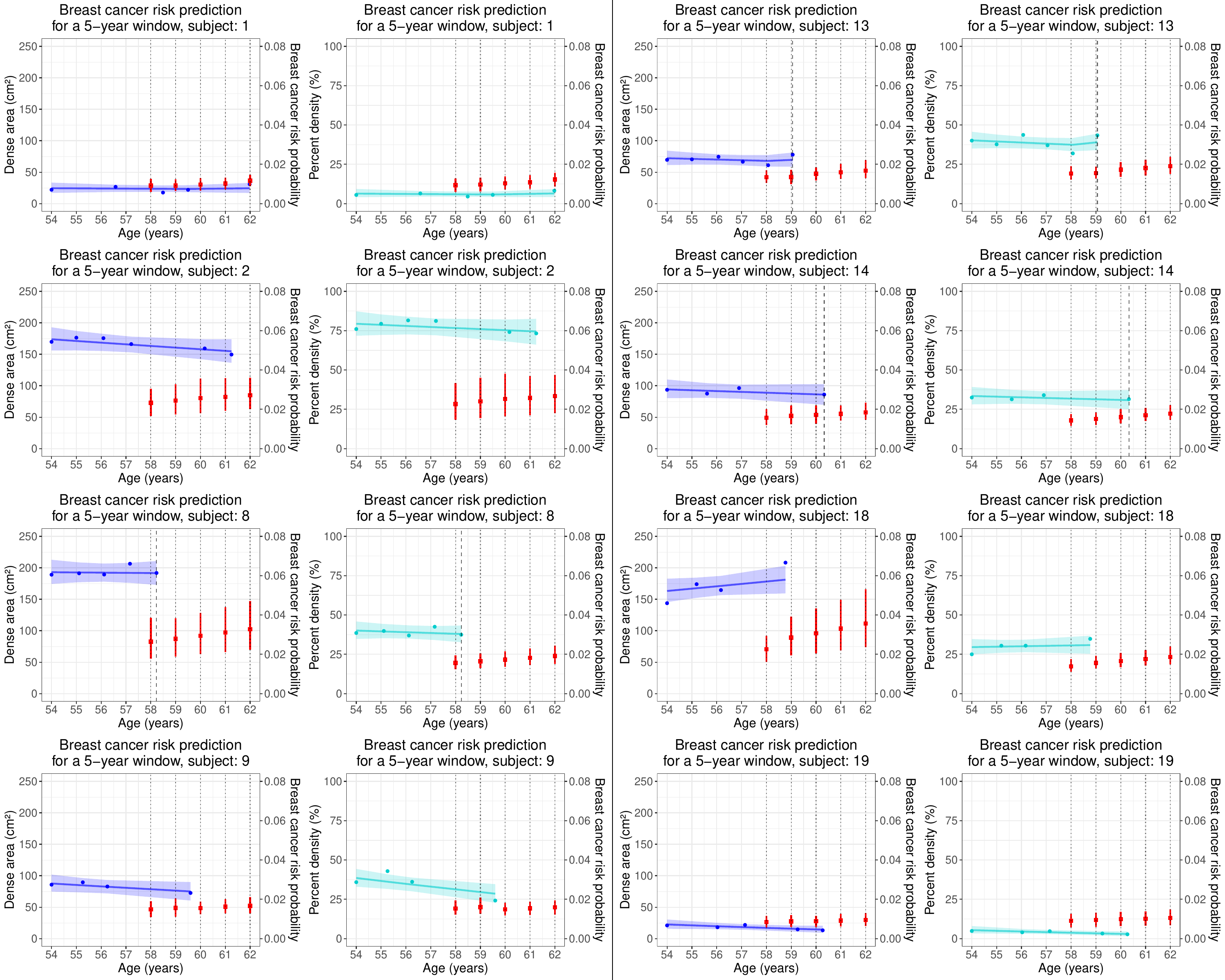}}
\caption{\textbf{Individual risk predictions in the next five years at landmark times ranging from \boldsymbol{$58$} to \boldsymbol{$62$} years for eight randomly selected women starting screening at the age of \boldsymbol{$54$} years.} Predictions were generated using the optimal joint model with either dense area (in blue) or percent density (in cyan) as the longitudinal biomarker. Blue and cyan dots represent the observed dense area and percent density evaluations, respectively. Blue and cyan bold lines illustrate the joint model's subject-specific longitudinal profiles for dense area and percent density, respectively, along with their corresponding 95\% credible intervals. Black dotted lines indicate landmark times, while black dashed lines represent breast cancer occurrence, if any. Red squares and bands denote the mean 5-year risk probabilities and their 95\% credible intervals. These predictions are based on the biomarker's evaluations up to a given landmark time, excluding subsequent observations.}
\label{f:dyn_pred_age54_w5}
\end{figure}

\begin{figure}[htbp]
 \centerline{\includegraphics[width=6in]{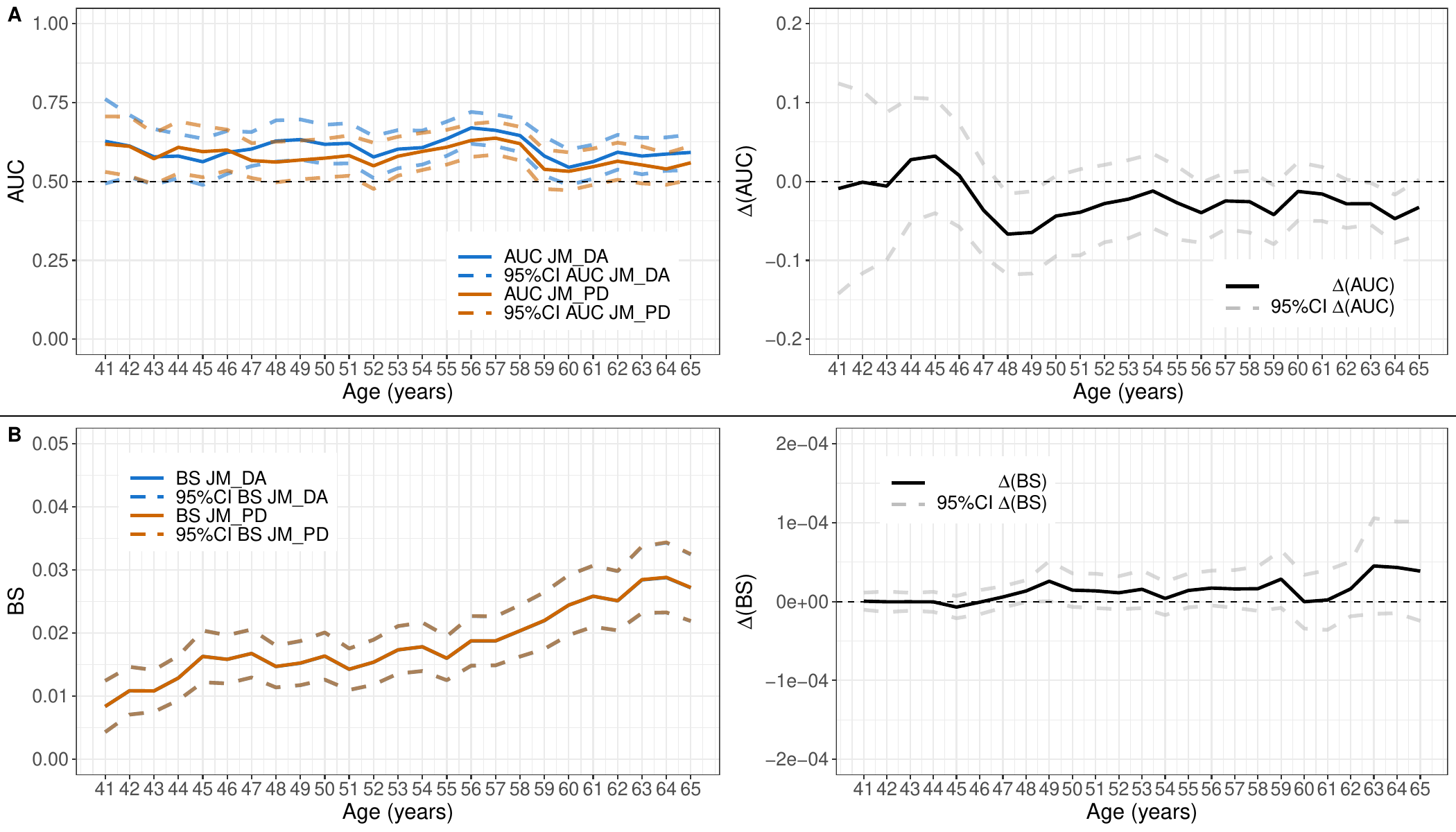}}
\caption{\textbf{Predictive accuracy of the optimal joint model (JM) with either dense area (DA) or percent density (PD) as the longitudinal biomarker within time interval $(s, s + w)$ when $s =\{41, 42, ..., 65\}$ and $w = 5$ years.} $\Delta(AUC)$ and $\Delta(BS)$ denote the differences in AUC and BS between JM with PD versus JM with DA. Dashed lines represent $95\%$ point-wise confidence intervals (CIs).}
\label{f:pred_accuracy_w5}
\end{figure}

Next, we assessed the predictive performance of Model 3 (Table~\ref{t:JM_DA_PD}) when considering dense area or percent density as the longitudinal biomarker using AUC and BS. Evaluation considered breast cancer risk predictions at landmark times $s =\{41, 42, ..., 65\}$ and a fixed prediction window of $w = 5$ years. Comparing the two joint models, we reported the $\Delta(AUC)$ and $\Delta(BS)$, representing the differences in AUC and BS between the model with percent density versus the one with dense area. Results in Figure~\ref{f:pred_accuracy_w5} suggested similar discrimination ability with a slight advantage for the joint model with dense area, though statistically insignificant as 0 was into the $95\%$ confidence interval of $\Delta(AUC)$. Models' calibration was comparable, with $\Delta(BS)$ values close to zero. Similar trends were observed for a fixed prediction window of $w = 2$ years (Appendix).

\section{Discussion}
\label{sec::discuss}
In this study, we introduced the DeepJoint algorithm, an open-source tool specially developed to analyze breast cancer screening data in average-risk women aged 40 and older. By combining a new deep learning model for quantitative mammographic density estimation with joint modeling, our approach captures the longitudinal relationship between density metrics and breast cancer risk, enabling dynamic and personalized risk predictions. We incorporated Bayesian inference and the Consensus Monte Carlo algorithm\cite{afonso_efficiently_2023} to handle large-scale screening datasets efficiently, making the DeepJoint algorithm a solution for practical, real-world clinical settings.

Building upon the MammoFL model,\cite{muthukrishnan_mammofl_2023} our deep learning model was adapted to analyze processed mammograms from multiple manufacturers, ensuring compatibility with diverse clinical settings. In addition to percent density, our deep learning model estimates dense area, addressing both metrics known to be independent risk factors for breast cancer. We implemented the model using PyTorch Lightning,\cite{falcon_pytorch_2019} which ensures ease of use and integration for a broader developer community. Training was conducted on 832 images from 208 women with available reference segmentation masks for both breast and dense tissue. We leveraged a pre-trained model on a large, multi-institutional dataset and implemented a 10-fold cross-validation process to ensure robustness. The model achieved high segmentation performance, and qualitative inspection (Appendix Figure~\ref{f:example_dl_perf}) confirmed good delineation of breast and dense regions. Although the training dataset was limited due to the labor-intensive manual creation of reference masks, the final model delivered visit-level mammographic density measurements aligned with BI-RADS assessments (Appendix Table~\ref{t:PD_&_DA_BI-RADS}). This supports the potential of our model to estimate density in real-world clinical applications reliably.

Then, we evaluated four joint models, each defined by a different link function associating the evolution of the longitudinal biomarker with breast cancer risk. This evaluation was conducted on a large dataset of 77,298 women, a scale that poses significant computational challenges for joint modeling. We utilized Bayesian inference and the Consensus Monte Carlo algorithm\cite{afonso_efficiently_2023} to address this, enabling efficient model fitting on such a large dataset. The best-performing model, regardless of whether dense area or percent density was used, incorporated both the current value and the slope of the longitudinal biomarker at a given age. Notably, this model revealed a positive association between the age at first screening and the longitudinal dense area, suggesting that women who begin screening at an older age tend to have higher mammographic density. However, this pattern was not observed with percent density, potentially suggesting an underlying selection bias within the population or other unmeasured factors influencing this difference.

Our analysis also showed a positive association between both the current level and the slope of the longitudinal biomarkers and breast cancer risk. These findings suggest that women with higher mammographic density are at an increased risk of breast cancer, and that those whose density declines more slowly with age may face a higher risk compared to those whose density decreases more rapidly. This observation aligns with previous findings, such as those by Ghosh et al. (2010),\cite{ghosh_independent_2010} which reported that the absence of lobular involution, an age-related reduction in breast lobules, is linked to higher breast cancer risk, particularly when combined with high mammographic density. Our results also generally concur with the findings of Illipse et al. (2023).\cite{illipse_studying_2023} However, there was a notable difference in the association between the slope of dense area and breast cancer risk. This discrepancy may stem from variations in the definition of the longitudinal biomarker, differences in covariate adjustments, or cohort characteristics between the two studies.

While our primary strategy was to model each biomarker separately to disentangle their individual contributions, we also conducted an exploratory analysis fitting a multivariate joint model on a reduced dataset of 2,501 women. The results were largely consistent with those of the univariate models. Nevertheless, given the known correlation between dense area and percent density, joint modeling of both metrcis remains an important direction for future work.

Besides, it is essential to note that although our dataset is large, it was sourced from only two centers and two manufacturers. This may limit the variability captured by the model and represent a potential generalizability constraint. Future work should include data from additional centers and vendors to better reflect real-world screening populations.

We also derived individualized and dynamic breast cancer risk probabilities over 2- and 5-year windows, noting differences in risk estimates depending on whether dense area or percent density was used as the longitudinal biomarker. These variations suggest that each metric captures distinct aspects of breast cancer risk, as previously discussed in the literature.\cite{baglietto_associations_2014, haars_measurements_2005} While our results highlight these differences, it is essential to recognize that the dataset used in this analysis lacks key adjustment variables, such as body mass index (BMI), which is critical for accurately modeling the relationship between mammographic density and breast cancer risk. While adjusting for BMI is crucial when considering percent density, it appears less impactful when using dense area.\cite{baglietto_associations_2014} 

In light of these results, it is important to precise that our objective in this study was two-fold: to explore the association between the longitudinal evolution of mammographic density and breast cancer risk, and to generate individualized, dynamic risk predictions. From an inferential standpoint, we used the joint model to examine how density trajectories, measured as either dense area or percent density, relate to the time-to-breast cancer event. However, we acknowledge that the absence of important clinical covariates such as BMI, menopausal status, and family history may lead to model misspecification. As emphasized in prior studies,\cite{heinze_variable_2018, wallisch_selection_2021} clinically relevant adjustment factors are necessary for deriving unbiased and interpretable estimates. As such, all association-related findings should be interpreted with appropriate caution.

In addition, considering recommendations from recent studies,\cite{calster_calibration_2019, Calster_performance_2024} we acknowledge that the Brier Score reflects both calibration and discrimination. To specifically assess calibration, we produced calibration plots using quantile-based grouping.\cite{riskRegression_gerds_2025} These plots revealed moderate miscalibration, likely due to missing prognostic variables. This diagnostic step points to the need for model refinement, possibly through recalibration or incorporation of additional clinical information.

Despite these limitations, the flexibility of our joint model allows for easy integration of additional covariates should these data become available. The model's structure readily accommodates both baseline and time-dependent adjustment variables, enabling more accurate and personalized risk assessments in future applications. Additionally, the model’s capacity to define variable prediction windows, from short-term to long-term risk predictions, offers adaptability for various clinical purposes, whether for immediate risk monitoring or extended surveillance. This flexibility is a strength, aligning with the growing emphasis on personalized screening strategies.\cite{roux_study_2022}

A key direction for future work is the external validation of the DeepJoint algorithm on independent datasets. Although not conducted in the current study, such validation is essential to assess generalizability across diverse populations, imaging platforms, and clinical settings. Publicly available datasets, such as the Emory Breast Imaging Dataset,\cite{jeong_theemory_2023} may be a first validation platform to benchmark our method against others and facilitate reproducibility. Such efforts will be critical to supporting clinical translation and real-world applicability.

Beyond mammographic density, additional mammographic texture features hold promise in refining breast cancer risk estimation. Jiang et al. (2023)\cite{jiang_supervised_2023} underscored the importance of incorporating image-based features alongside mammographic density for enhanced breast cancer prediction, proposing robust dimension reduction techniques for imaging data in time-to-event analyses. We plan to integrate these image-based features into the joint model framework in future work. We aim to evaluate how these additional features and the longitudinal evolution of mammographic density influence breast cancer risk.

In addition to FFDM, digital breast tomosynthesis (DBT) has shown potential in improving breast cancer detection in the screening setting. DBT generates 2D synthetic mammography (2DSM) images by reconstructing tomosynthesis slices, avoiding additional radiation exposure, which is especially beneficial for women participating in routine screening. Studies have demonstrated high agreement between mammographic density assessments from FFDM and 2DSM images.\cite{conant_agreement_2017, haider_comparison_2018, moshina_comparing_2020} We intend to extend our method's application to 2DSM images, broadening its utility across different imaging modalities.

Lastly, Dadsetan et al. (2022)\cite{dadsetan_deep_2022} explored a full deep learning methodology for breast cancer assessment by capturing spatiotemporal changes in bilateral breast tissue features. While this approach holds promise, multiple challenges constrain its practical implementation, including the necessity for a substantial dataset, incompatibility with varying visit numbers and intervals, and limited interpretability attributed to the "black-box" nature. In contrast, our approach combines advanced deep learning for density estimation with joint modeling, ensuring both accurate risk predictions and clear interpretability. This integration balances complexity and transparency, representing a key innovation in our work and making our model a robust and practical tool for clinical application.
\section*{Acknowledgement}
The authors gratefully acknowledge the "Institut National du Cancer" (INCa\textunderscore16049) and "La Ligue de la Gironde et des Landres" for funding this research. Additional funding was received through the PIA3 (Investment for the Future--project reference 17-EURE-0019).

\newpage

\section*{Appendix}
\renewcommand{\thefigure}{A.\arabic{figure}}
\setcounter{figure}{0} 
\renewcommand{\thetable}{A.\arabic{table}}
\setcounter{table}{0} 
\subsection*{A. \enspace The joint model estimation}
Let $\boldsymbol{\theta}=(\boldsymbol{\beta}, \boldsymbol{\gamma}, \boldsymbol{\alpha}, \sigma_{\epsilon}^{2}, \boldsymbol{B}, h_{0}(t))$. The baseline hazard function $h_{0}(t)$ is usually approximated when using joint models to facilitate likelihood inference. Here, we model the baseline hazard function using a penalised B-spline, where $log(h_{0}(t)) = \sum_{q=1}^{Q} \boldsymbol{\gamma_{h_{0},q}} B_{q}(t, \upsilon)$,
with $B_{q}(t, \upsilon)$ denoting the q-th basis function of a B-spline with knots $\upsilon_{1}, ..,\upsilon_{Q}$ and $\boldsymbol{\gamma}_{h_{0}}$ the vector of spline coefficients.\cite{rizopoulos_jmbayes2_2023}

Let $L_{i}(\boldsymbol{\theta})$ be each individual contribution to the likelihood. The latter is expressed under the assumption of the conditional independence between the biomarker $\boldsymbol{Y_{i}}$ and the event time $T_{i}$ given the random effects $\boldsymbol{b_{i}}$ and is defined as follows
\begin{equation} \label{eq:indlikelihood_surv}
	\begin{array}{lll}
		L_{i}(\boldsymbol{\theta}) & = & f_{\boldsymbol{Y_{i}}, (T_{i}, \delta_{i})}(\boldsymbol{Y_{i}}, (T_{i}, \delta_{i});\boldsymbol{\theta}) \\[0.1cm]
		L_{i}(\boldsymbol{\theta}) & = & \mathlarger{\int_{\boldsymbol{b_{i}}}} \ \mathlarger{\prod_{j=1}^{n_{i}}} \ f_{Y_{i}|\boldsymbol{b_{i}}} (Y_{ij} | \boldsymbol{b_{i}}; \boldsymbol{\theta}) \ S_{i}(T_{i} | \boldsymbol{b_{i}}; \boldsymbol{\theta}) \ h_{i}(T_{i} | \boldsymbol{b_{i}}; \boldsymbol{\theta})^{\delta_{i}} \ f_{\boldsymbol{b_{i}}}(b; \boldsymbol{\theta}) \ d\boldsymbol{b_{i}}\\[0.2cm]
		& = & \mathlarger{\int_{\boldsymbol{b_{i}}}} \ \frac{1}{(\sqrt{2 \pi \sigma_{\epsilon}^{2}})^{n_{i}}} \ \mathlarger{\prod_{j=1}^{n_{i}}} \exp \Big{(}-\frac{(Y_{ij} \ - \ \boldsymbol{X_{ij}}^{D^\top} \ \boldsymbol{\beta} \ + \ \boldsymbol{Z_{ij}}^{D^\top} \ \boldsymbol{b_{i}})^{2}}{2\sigma_{\epsilon}^{2}}\Big{)} \\
		& & \times \ h_{i}(T_{i} | \boldsymbol{b_{i}}; \boldsymbol{\theta})^{\delta_{i}} \exp \Big{(} -\mathlarger{\int_{t_{0i}}^{T_{i}}} h_{i}(s|\boldsymbol{b_{i}})ds)\Big{)} \times f_{\boldsymbol{b_{i}}}(b; \boldsymbol{\theta}) \ d\boldsymbol{b_{i}}\\[0.2cm]
	\end{array}
\end{equation}
with the density function of random effects $f_{\boldsymbol{b_{i}}}(b; \boldsymbol{\theta})$ is assumed to follow a normal distribution with mean zero and variance-covariance matrix $\boldsymbol{B}$.
\newpage
\subsection*{B. \enspace Predictive accuracy measures}
\paragraph{B.1 \enspace Dynamic AUC}
The AUC assesses the discrimination ability of a predictive tool, with higher values indicating the tool's capacity to give higher predicted risks of event for subjects who are more likely to experience the event than for subjects who are less likely to experience it. Let $D(s,w)$ be the observed outcome in a specific time span $[s, s+w]$ in women still at risk at time $s$. Given that $D(s,w)$ is not always observable in the presence of right-censored data, we define $\tilde{D}(s,w) = \mathds{1}_{s < T < s+w}$, that equals $1$ when the event of interest occurs within $[s, s+w]$, and $0$ otherwise. To handle right-censored data, the dynamic AUC is estimated using the the Inverse Probability of Censoring Weighting (IPCW) method.\cite{blanche_quantifying_2015} As a result, the expected AUC at a landmark time $s$ for a fixed prediction window $w$ is given by
\begin{equation}
\label{eq:AUC_surv}
	\begin{array}{ll}
	    \displaystyle 
	    AUC(s,w) & = \mathrm{P}\Big{(}\pi_{i}(s+w) > \pi_{j}^{*}(s+w) | D_{i}(s,w)=1, \ D_{j}(s,w)=0, \ T_{i}>s, \ T_{j}>s\Big{)}\\[0.3cm]
	    & = \frac{\sum_{i=1}^{N} \sum_{j=1}^{N} \ \mathds{1}_{(\pi_{i}(s+w)>\pi_{j}^{*}(s+w))} \ \tilde{D_{i}}(s,w) \ (1 - \tilde{D_{j}}(s,w)) \ \widehat{W_{i}}(s,w) \ \widehat{W_{j}}(s,w)}{\sum_{i=1}^{N} \sum_{j=1}^{N} \ \tilde{D_{i}}(s,w) \ (1-\tilde{D_{j}}(s,w)) \ \widehat{W_{i}}(s,w) \ \widehat{W_{j}}(s,w)}
	\end{array}
\end{equation}

here, $\widehat{W_{i}}(s,w)$ are weights to account for right-censoring, defined as
\begin{equation} \label{eq:WeightBS_surv}
	\begin{array}{ll}
	    \displaystyle 
	    \widehat{W_{i}}(s,w) & = \frac{\mathds{1}_{(T_{i} > s+w)}}{\tilde{G}(s+w|s)} + \frac{\mathds{1}_{(s < T_{i}<s+w)}\delta_{i}}{\tilde{G}(T_{i}|s)}\\
	\end{array}
\end{equation}

with $\hat{G(u)}$ as the Kaplan-Meier estimator of the survival function at censoring time $u$, such as $\forall u > s$, $\hat{G}(u|s)=\hat{G}(u)/\hat{G}(s)$.

\paragraph{B.2 \enspace Dynamic Brier score}
The Brier score represents the mean squared error between predicted probabilities $\pi(s+w)$ at a given time $s+w$ and the observed outcome $D(s,w)$ in women still at risk at time $s$. Consequently, the lower and closest to zero is the BS, the better. Dynamic BS, accounting for right-censoring, is estimated using the IPCW method.\cite{blanche_quantifying_2015} The expected BS at a horizon time $s+w$ is defined as

\begin{equation}
\label{eq:BS_surv}
	\begin{array}{ll}
	    \displaystyle 
	    BS(s,w) & = \mathit{E}\Big{[}(D(s,w) - \pi_{i}(s+w))^{2}| T > s \Big{]}\\
	    & = \frac{1}{\sum_{i=1}^{N} \mathds{1}_{T_{i}>s}} \sum_{i=1}^{N} \widehat{W_{i}}(s,w)\Big{(}\tilde{D_{i}}(s,w) - \pi_{i}(s+w)\Big{)}^2 \\
	\end{array}
\end{equation}
\newpage
\subsection*{C. \enspace Data and code}
All analyses in this work were conducted using the code available in the \href{https://github.com/manelrakez/deepjoint-algo}{DeepJoint GitHub repository}. The DeepJoint algorithm integrates deep learning for quantitative mammographic density estimation with joint modeling to explore the longitudinal relationship between mammographic density and breast cancer risk. The deep learning model is implemented in Python, while the joint model is written in R.

Below, we provide a step-by-step example of how to fit the joint model (Model 3 from the main manuscript) using the Monte Carlo consensus method. To perform this analysis, two datasets are required: First, the longitudinal dataset (\texttt{data}) that contains repeated measurements of mammographic density and covariates. Second the survival dataset (\texttt{data\_surv}) including information on breast cancer events.

Tables~\ref{t:extract_data} and \ref{t:extract_data_surv} provide extracts from these datasets. In Table~\ref{t:extract_data}, the variable "pts\_age\_modif" represents the woman's age at each visit (in years), adjusted by subtracting 40 to account for left truncation starting at age 40. The variables "sqrt\_da\_cm2" and "sqrt\_pd" correspond to the longitudinal biomarkers for dense area (cm²) and percent density (\%), respectively.
In Table~\ref{t:extract_data_surv}, the variable "start" refers to the woman’s age (adjusted as described above) at her first mammogram. The variable "end" represents her age at the last mammography visit either preceding a positive biopsy or before censoring (if no event occurred).
\begin{table}[htbp]
\caption{\textbf{Extract from dataset \texttt{data} for the longitudinal measurements and covariates}}
\label{t:extract_data}
\begin{center}
\begin{small}
\begin{tabular}{cccccccc}
\hline
pts\_id & bl\_age & pts\_age & pts\_age\_modif & sqrt\_da\_cm2 & sqrt\_pd & manuf & event\\
\hline
1 & 48 & 48.00 & 8.00 & 9.35 & 7.28 & A & 0\\
1 & 48 & 49.43 & 9.43 & 9.30 & 7.06 & A & 0\\
2 & 63 & 63.00 & 23.00 & 10.08 & 6.79 & A & 0\\
2 & 63 & 64.21 & 24.21 & 9.53 & 6.48 & A & 0\\
2 & 63 & 65.55 & 25.55 & 9.70 & 6.40 & A & 0\\
2 & 63 & 66.54 & 26.54 & 9.44 & 6.51 & A & 0\\
\vdots & \vdots & \vdots & \vdots & \vdots & \vdots & \vdots & \vdots\\
\hline
\end{tabular}
\end{small}
\end{center}
\end{table}
\begin{table}[htbp]
\caption{\textbf{Extract from dataset \texttt{data\_surv} for breast cancer event and covariates}}
\label{t:extract_data_surv}
\begin{center}
\begin{small}
\begin{tabular}{cccc}
\hline
pts\_id & start & end & event\\
\hline
1 & 8 & 9.43 & 0\\
2 & 23 & 27.54 & 0\\
3 & 34 & 36.05 & 0\\
4 & 16 & 18.43 & 0\\
\vdots & \vdots & \vdots & \vdots\\
\hline
\end{tabular}
\end{small}
\end{center}
\end{table}

\clearpage
Here is how to perform the estimation of the joint model on a subset of 2,501 randomly selected women. Results are only for illustration purposes and should not be interpreted.
\begin{lstlisting}[style=Rstyle]
#Data and src codes are available on https://github.com/manelrakez/deepjoint-algo
source("~/deepjoint-algo/src/deepjoint_r/jm_consensus_source_script_CV.R")
pcks <- c("JMbayes2")
invisible(lapply(pcks, require, character.only = TRUE)) #Load packages
directory <- "~/deepjoint-algo/test_data/JM_test"
setwd(directory) #Define your directory

# Import datasets
data <- read.csv2("data.csv")
data_surv <- read.csv2("data_surv.csv")

# Example for Model 3 with dense area as the longitudinal biomarker and CVCS as the link function. For Model 3 with percent density replace "sqrt_da_cm2" by "sqrt_pd"
# 1/ Model specification
n_slices <- 3
n_chains <- 3
n_iter <- 3500L
n_burnin <- 500L
n_iter_net <- n_iter-n_burnin
args_long <- list(fixed = sqrt_da_cm2 ~ pts_age_modif + bl_age + manuf, 
                  random = ~ pts_age_modif | pts_id, #Random intercept and slope
                  control = lmeControl(opt = "optim"))
args_surv <- list(formula = survival::Surv(start, end, event) ~ 1)

args_jm_CVCS <- list(time_var = "pts_age_modif", 
                       n_iter = n_iter, n_burnin = n_burnin, n_thin = 1L, 
                       parallel = 'multicore',
                       functional_forms = ~ value(sqrt_da_cm2) + slope(sqrt_da_cm2))
                       
# 2/ Fit the model with parallel run using the consensus algorithm
mod_split_CVCS <- jm_parallel(args_jm_CVCS,
                              id_var="pts_id",
                              data_long = data,
                              data_surv = data_surv,
                              args_long = args_long,
                              args_surv = args_surv,
                              n_slices = n_slices,
                              CV_pred = 1, # 1 if you don't want to save split data
                              event_var = "event")
                              
# 3/ Create consensus model
jm_cons_CVCS <- jm_consensus(mod_split_CVCS$jm, combine_chains = FALSE)
\end{lstlisting}
  \newpage
\begin{lstlisting}[style=Rstyle]
# 4/Summarize model parameters in (mean, lower, and upper quantiles)
summary_mcmc <- function(mcmc_list, param_type) {
  summarize_mcmc <- function(x) {
    res_mean <- round(apply(x, 2, mean), 3)
    res_lower <- round(apply(x, 2, quantile, probs = 0.025), 3)
    res_upper <- round(apply(x, 2, quantile, probs = 0.975), 3)
    return(data.frame(mean = res_mean, lower_95CI = res_lower, upper_95CI = res_upper))
  }
  if (param_type == "betas") {
    res_param <- summarize_mcmc(mcmc_list$betas1)
  } else if (param_type == "sigmas") {
    res_param <- summarize_mcmc(mcmc_list$sigmas)
  } else if (param_type == "gammas") {
    res_param <- summarize_mcmc(mcmc_list$gammas)
  } else if (param_type == "alphas") {
    res_param <- summarize_mcmc(mcmc_list$alphas)
  } else {
    stop("Invalid parameter type. Please specify 'betas', 'sigmas', 'gammas', or 'alphas'.")
  }
  return(res_param)
}
# 4.a/ Calculate summary for 'betas'
betas_summary <- summary_mcmc(jm_cons_CVCS$mcmc, "betas")
# 4.b/ Calculate summary for 'sigmas'
sigmas_summary <- summary_mcmc(jm_cons_CVCS$mcmc, "sigmas")
# 4.c/ Calculate summary for 'alphas'
alphas_summary <- summary_mcmc(jm_cons_CVCS$mcmc, "alphas")
# 4.d/ All results
All_param_summary <- rbind(betas_summary, sigmas_summary, alphas_summary)
print(All_param_summary)
\end{lstlisting}

\begin{figure}[htbp]
\begin{footnotesize}
\centerline{\includegraphics[width=4in]{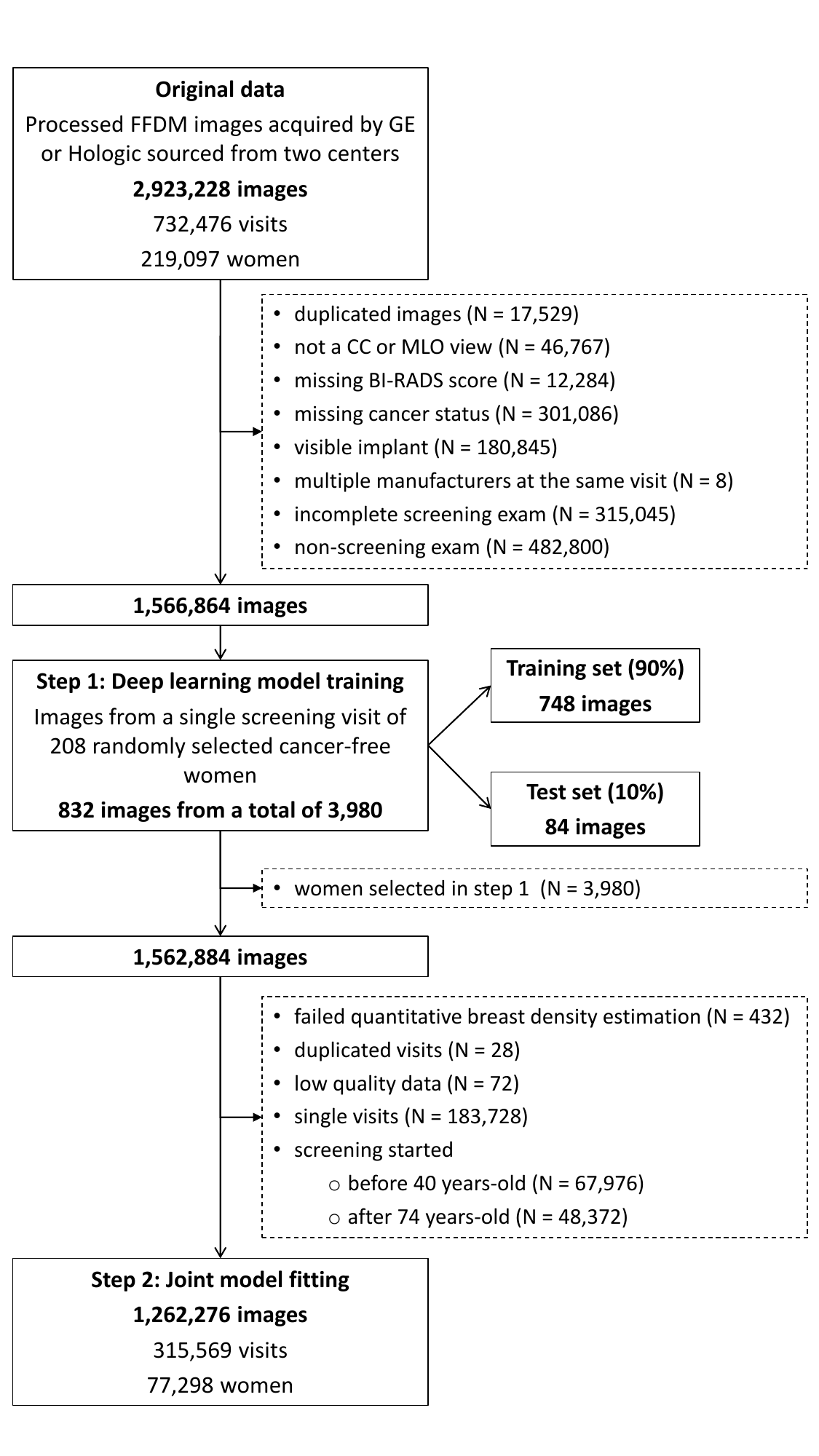}}
\caption{\textbf{Flowchart summarizing the data selection process for model development.} The original dataset consisted of over $2.9$ million processed full-field digital mammography (FFDM) images from $219,097$ women screened at two centers using GEHC or Hologic systems. After applying exclusion criteria, approximately $1.56$ million images remained. Step 1 describes the subset used for training and testing the deep learning segmentation model. Step 2 outlines the selection of the longitudinal cohort used for joint model fitting, resulting in a final study population of $77,298$ women contributing to over $315,000$ visits.}
\label{f:FlowChart}
\end{footnotesize}
\end{figure}

\begin{figure}[htbp]
 \centerline{\includegraphics[width=7in]{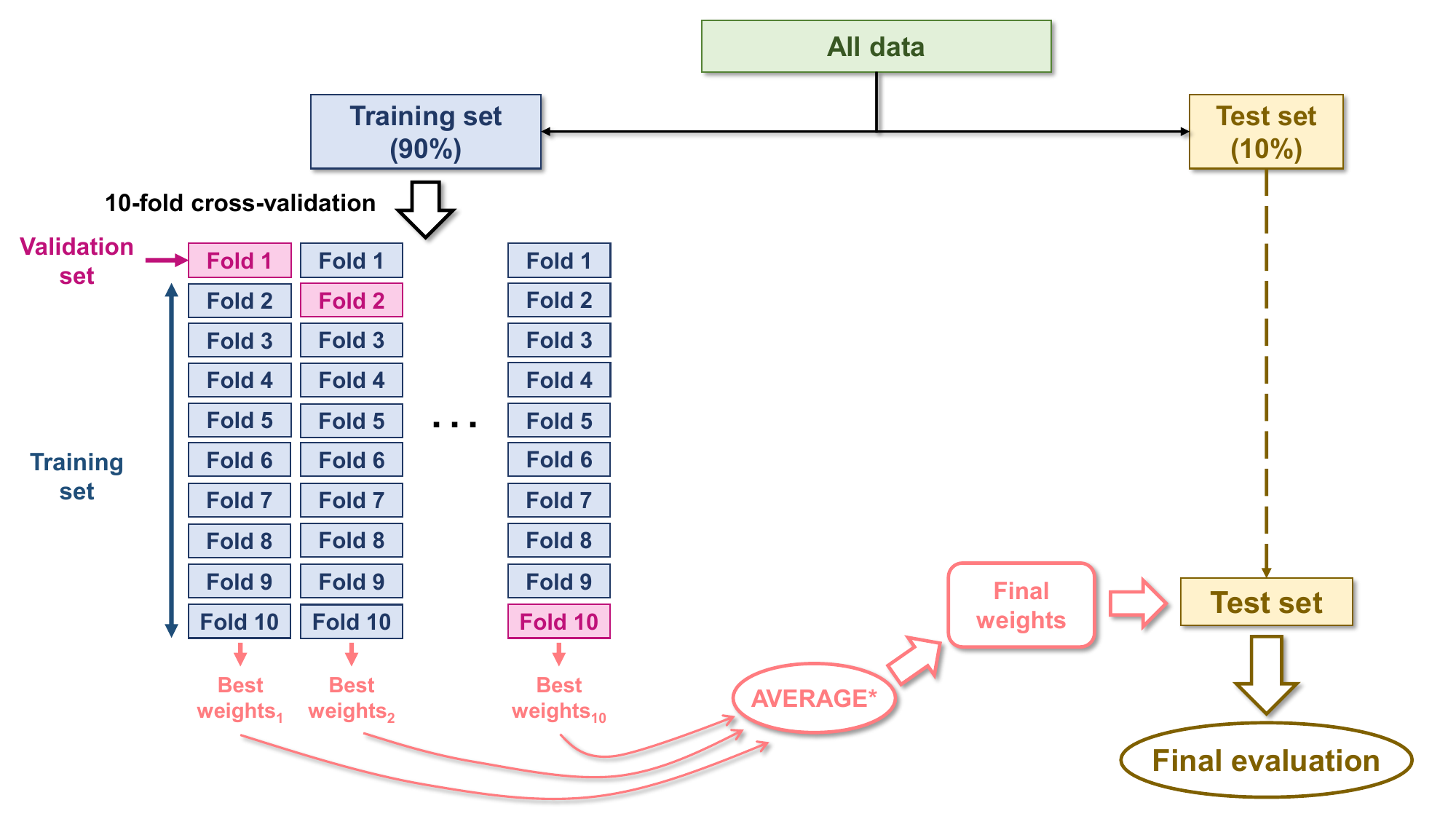}}
\caption{\textbf{Training process of the deep learning component of the DeepJoint algorithm using 10-fold cross-validation.} The model was trained using a 10-fold cross-validation scheme. Best model weights across folds were aggregated using a simple (unweighted) average, as indicated by the asterisk in the diagram.}
\label{f:CrossVal}
\end{figure}

\begin{figure}[htbp]
\begin{footnotesize}
\centerline{\includegraphics[width=4.5in]{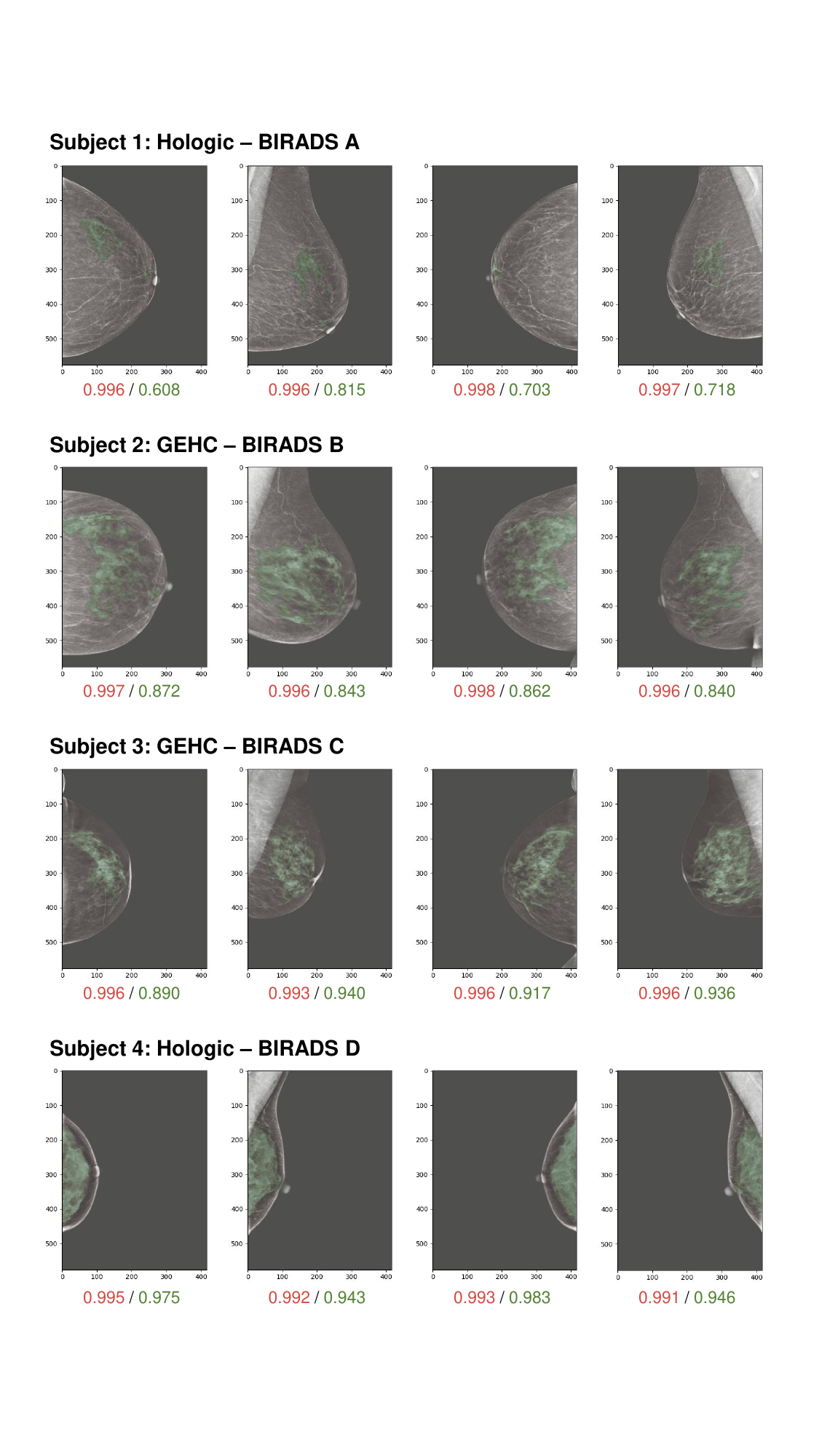}}
\caption{\textbf{Performance of the deep learning model on processed mammograms from the test set across different BI-RADS classes and two manufacturers.} The breast area is highlighted in red, while the dense area is shown in green. The Dice Similarity coefficient (DSC) for both breast and dense tissue segmentation are displayed beneath each mammogram.}
\label{f:example_dl_perf}
\end{footnotesize}
\end{figure}

\begin{figure}[htbp]
 \centerline{\includegraphics[width=6in]{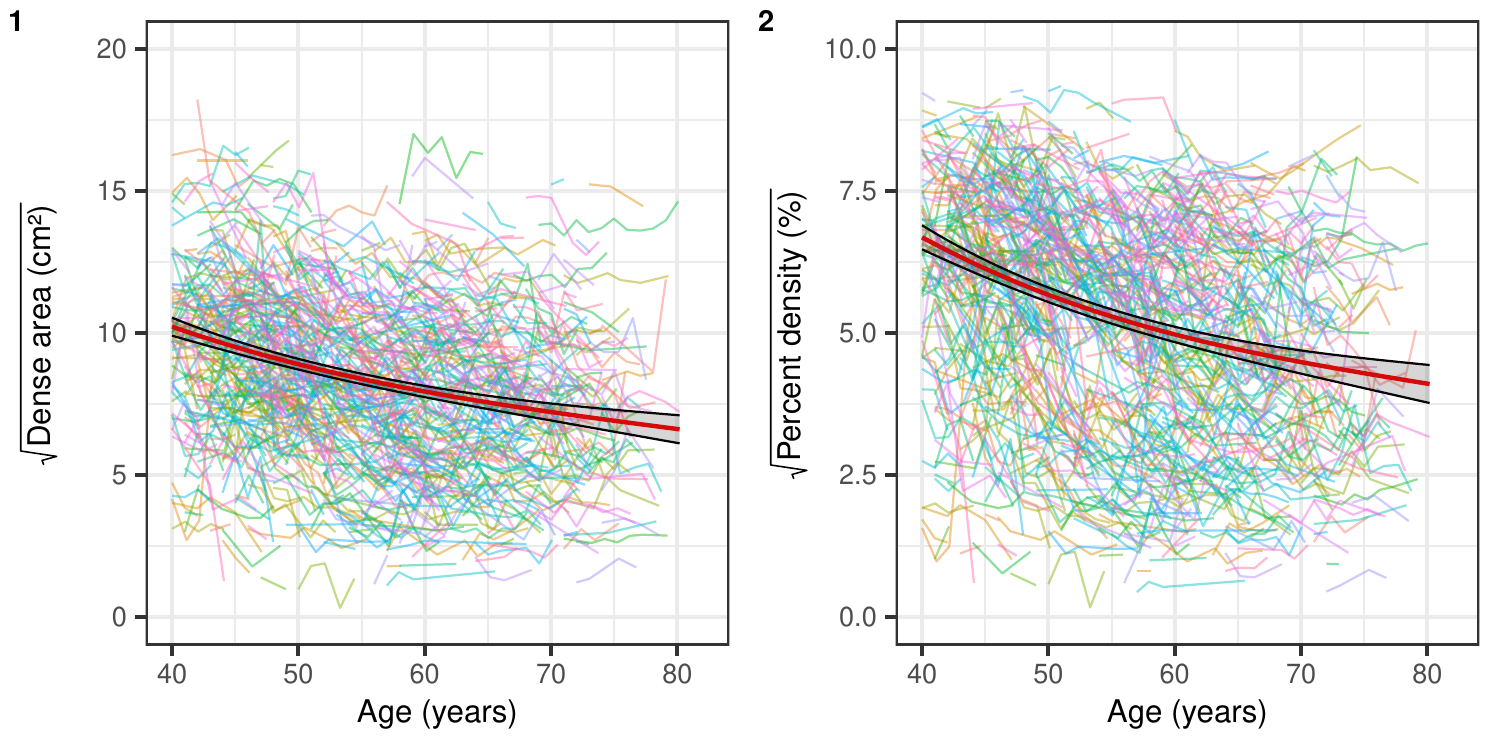}}
\caption{\textbf{Evolution of mammographic density metrics over age for 1,000 randomly selected women.} The square root of (1) dense area (in $cm^{2}$) and (2) percent density (in \%) are plotted against age. The smoothed average trend is shown in red, with 95\% confidence bands displayed in black.}
\label{f:Spaghetti}
\end{figure}

\begin{figure}[htbp]
 \centerline{\includegraphics[width=6.5in]{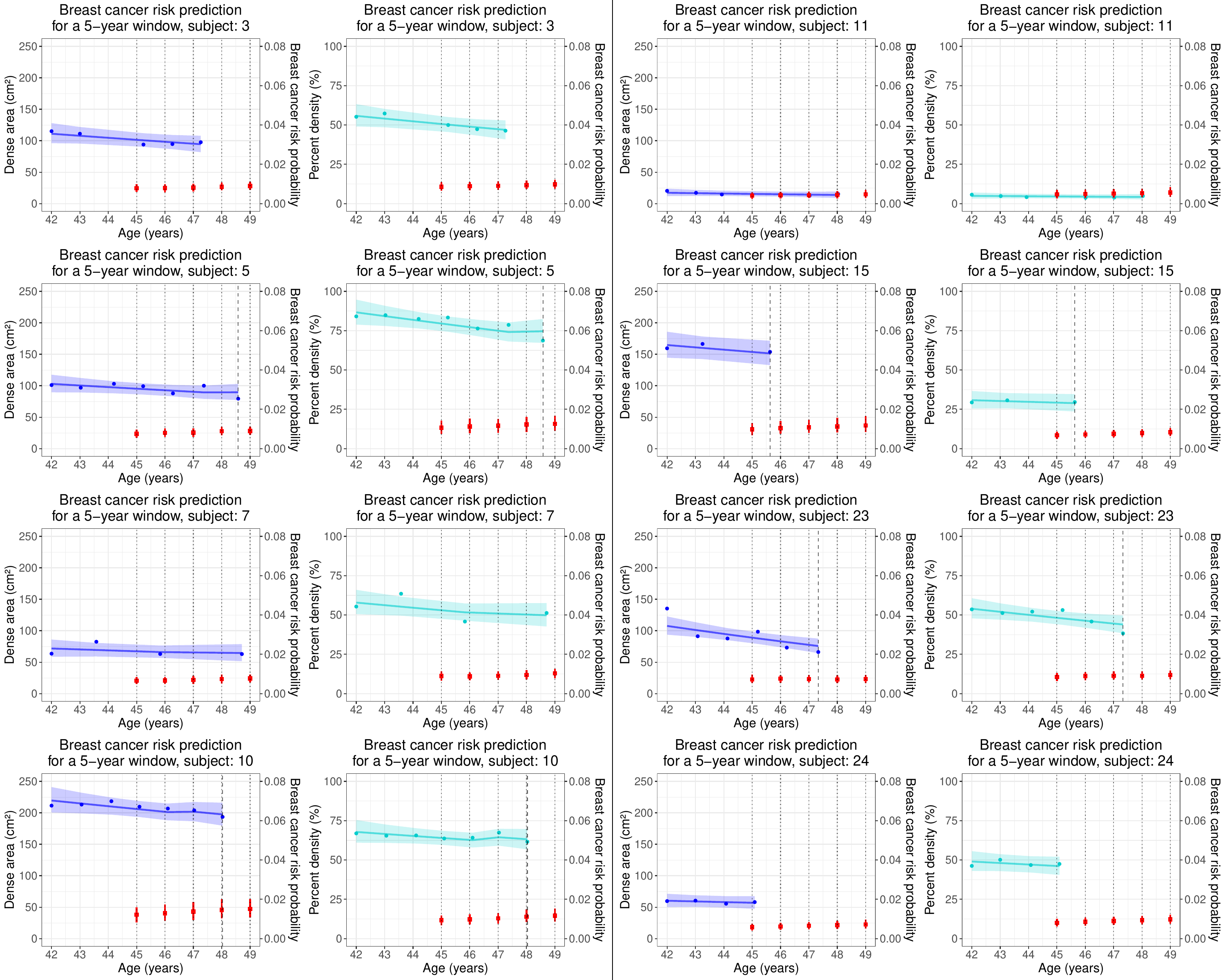}}
\caption{\textbf{Individual risk predictions in the next five years at landmark times ranging from $45$ to $49$ years for eight randomly selected women starting screening at the age of 42 years.} Predictions were generated using the optimal joint model with either dense area (in blue) or percent density (in cyan) as the longitudinal biomarker. Blue and cyan dots represent the observed dense area and percent density evaluations, respectively. Blue and cyan bold lines illustrate the joint model's subject-specific longitudinal profiles for dense area and percent density, respectively, along with their corresponding 95\% credible intervals. Black dotted lines indicate landmark times, while black dashed lines represent breast cancer occurrence, if any. Red squares and bands denote the mean 5-year risk probabilities and their 95\% credible intervals. These predictions are based on the biomarker's evaluations up to a given landmark time, excluding subsequent observations.}
\label{f:dyn_pred_age42_w5}
\end{figure}

\begin{figure}[htbp]
 \centerline{\includegraphics[width=6.5in]{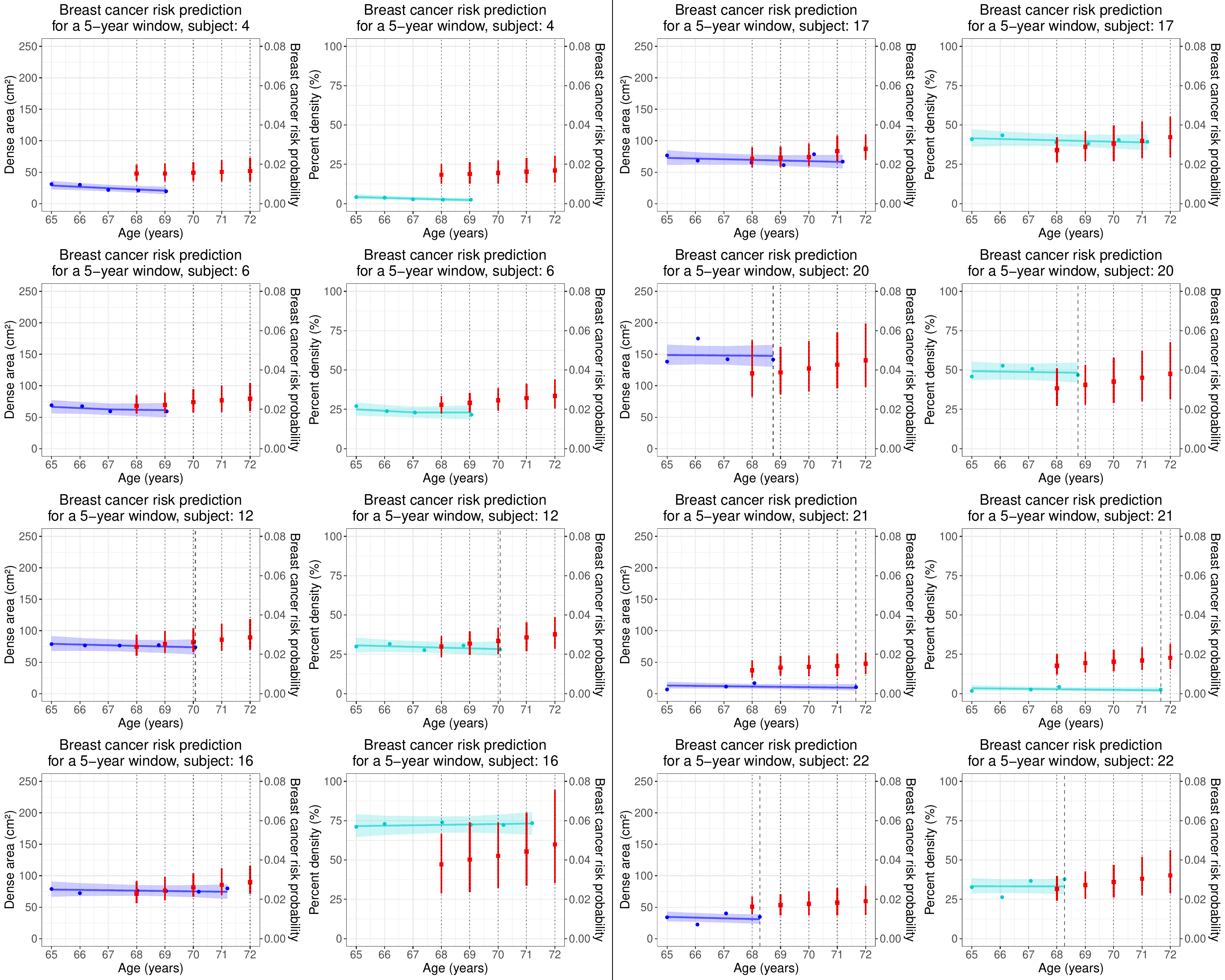}}
\caption{\textbf{Individual risk predictions in the next five years at landmark times ranging from $68$ to $72$ years for eight randomly selected women starting screening at the age of 65 years.} Predictions were generated using the optimal joint model with either dense area (in blue) or percent density (in cyan) as the longitudinal biomarker. Blue and cyan dots represent the observed dense area and percent density evaluations, respectively. Blue and cyan bold lines illustrate the joint model's subject-specific longitudinal profiles for dense area and percent density, respectively, along with their corresponding 95\% credible intervals. Black dotted lines indicate landmark times, while black dashed lines represent breast cancer occurrence, if any. Red squares and bands denote the mean 5-year risk probabilities and their 95\% credible intervals. These predictions are based on the biomarker's evaluations up to a given landmark time, excluding subsequent observations.}
\label{f:dyn_pred_age65_w5}
\end{figure}

\begin{figure}[htbp]
 \centerline{\includegraphics[width=6.5in]{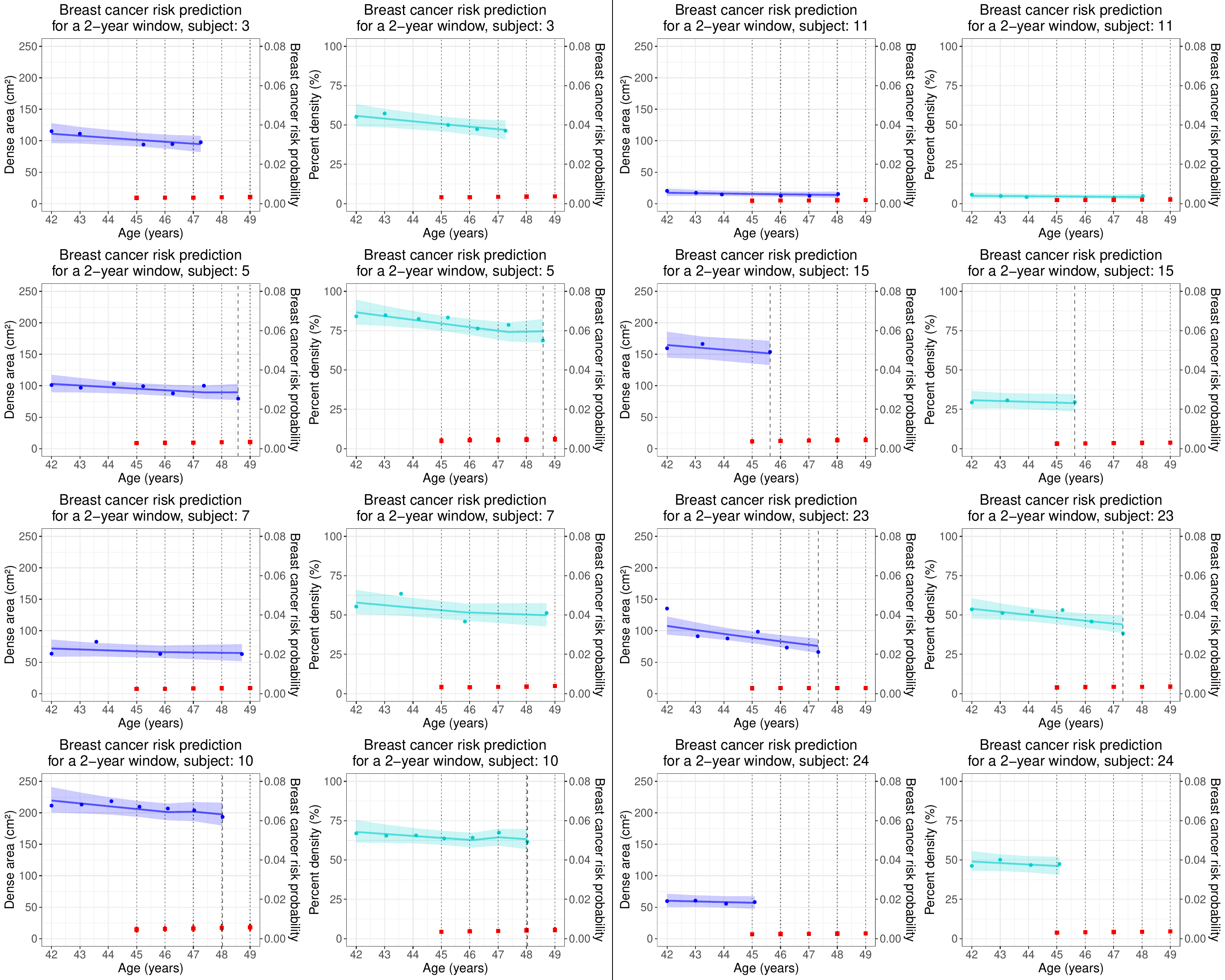}}
\caption{\textbf{Individual risk predictions in the next five years at landmark times ranging from $45$ to $49$ years for eight randomly selected women starting screening at the age of 42 years.} Predictions were generated using the optimal joint model with either dense area (in blue) or percent density (in cyan) as the longitudinal biomarker. Blue and cyan dots represent the observed dense area and percent density evaluations, respectively. Blue and cyan bold lines illustrate the joint model's subject-specific longitudinal profiles for dense area and percent density, respectively, along with their corresponding 95\% credible intervals. Black dotted lines indicate landmark times, while black dashed lines represent breast cancer occurrence, if any. Red squares and bands denote the mean 2-year risk probabilities and their 95\% credible intervals. These predictions are based on the biomarker's evaluations up to a given landmark time, excluding subsequent observations.}
\label{f:dyn_pred_age42_w2}
\end{figure}

\begin{figure}[htbp]
 \centerline{\includegraphics[width=6.5in]{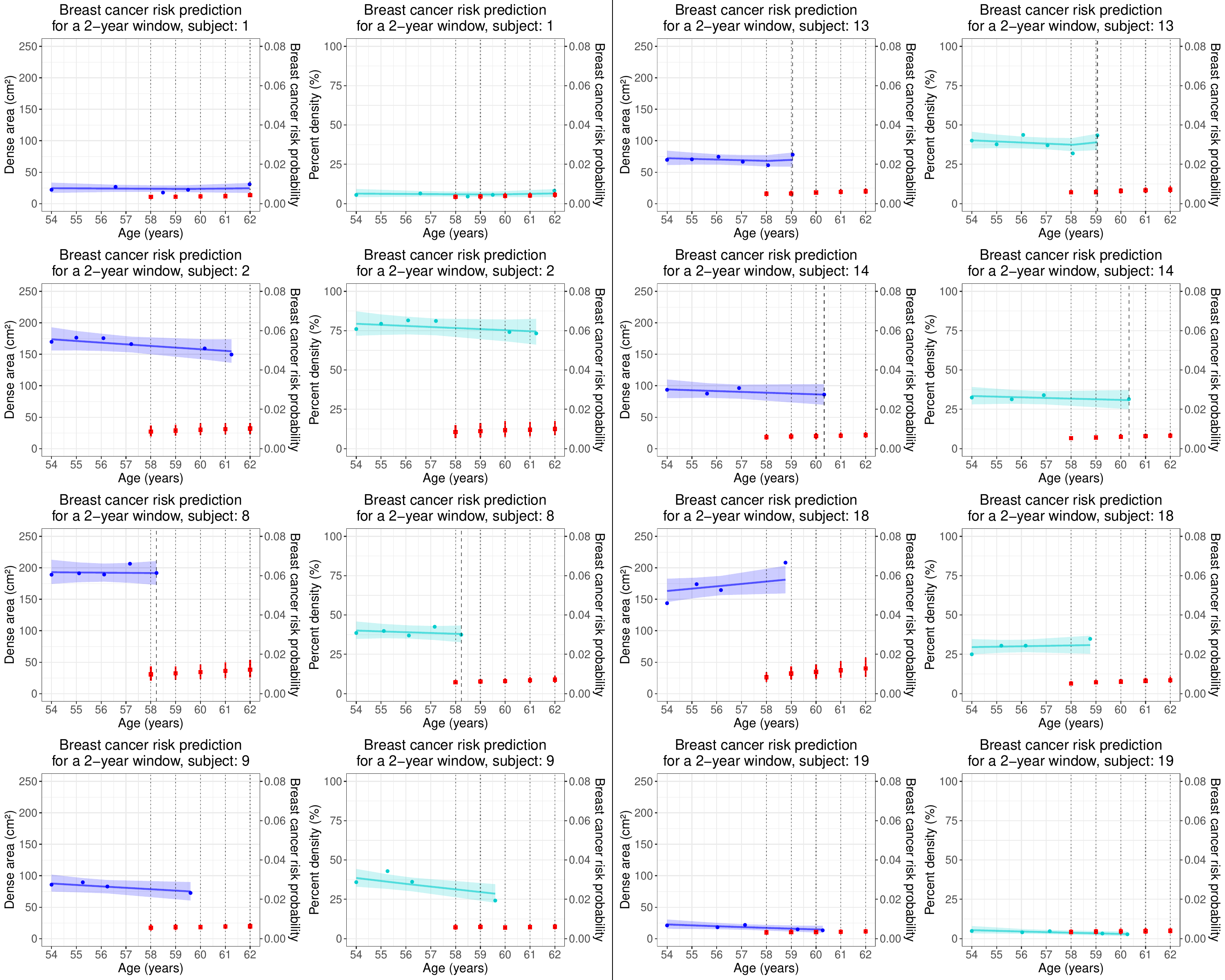}}
\caption{\textbf{Individual risk predictions in the next five years at landmark times ranging from $58$ to $62$ years for eight randomly selected women starting screening at the age of 54 years.} Predictions were generated using the optimal joint model with either dense area (in blue) or percent density (in cyan) as the longitudinal biomarker. Blue and cyan dots represent the observed dense area and percent density evaluations, respectively. Blue and cyan bold lines illustrate the joint model's subject-specific longitudinal profiles for dense area and percent density, respectively, along with their corresponding 95\% credible intervals. Black dotted lines indicate landmark times, while black dashed lines represent breast cancer occurrence, if any. Red squares and bands denote the mean 2-year risk probabilities and their 95\% credible intervals. These predictions are based on the biomarker's evaluations up to a given landmark time, excluding subsequent observations.}
\label{f:dyn_pred_age54_w2}
\end{figure}

\begin{figure}[htbp]
 \centerline{\includegraphics[width=6.5in]{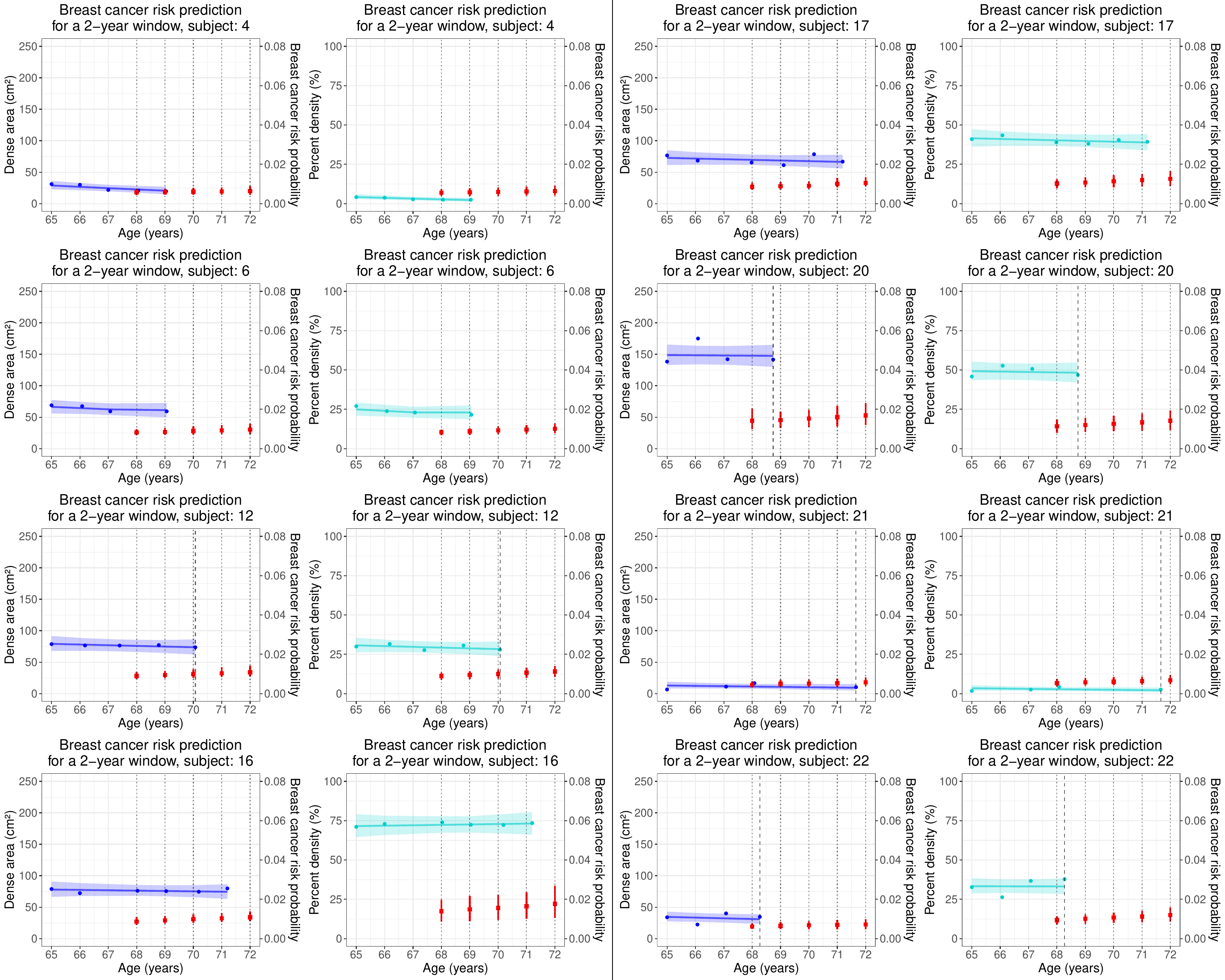}}
\caption{\textbf{Individual risk predictions in the next five years at landmark times ranging from$68$ to $72$ years for eight randomly selected women starting screening at the age of 65 years.} Predictions were generated using the optimal joint model with either dense area (in blue) or percent density (in cyan) as the longitudinal biomarker. Blue and cyan dots represent the observed dense area and percent density evaluations, respectively. Blue and cyan bold lines illustrate the joint model's subject-specific longitudinal profiles for dense area and percent density, respectively, along with their corresponding 95\% credible intervals. Black dotted lines indicate landmark times, while black dashed lines represent breast cancer occurrence, if any. Red squares and bands denote the mean 2-year risk probabilities and their 95\% credible intervals. These predictions are based on the biomarker's evaluations up to a given landmark time, excluding subsequent observations.}
\label{f:dyn_pred_age65_w2}
\end{figure}

\begin{figure}[htbp]
 \centerline{\includegraphics[width=6in]{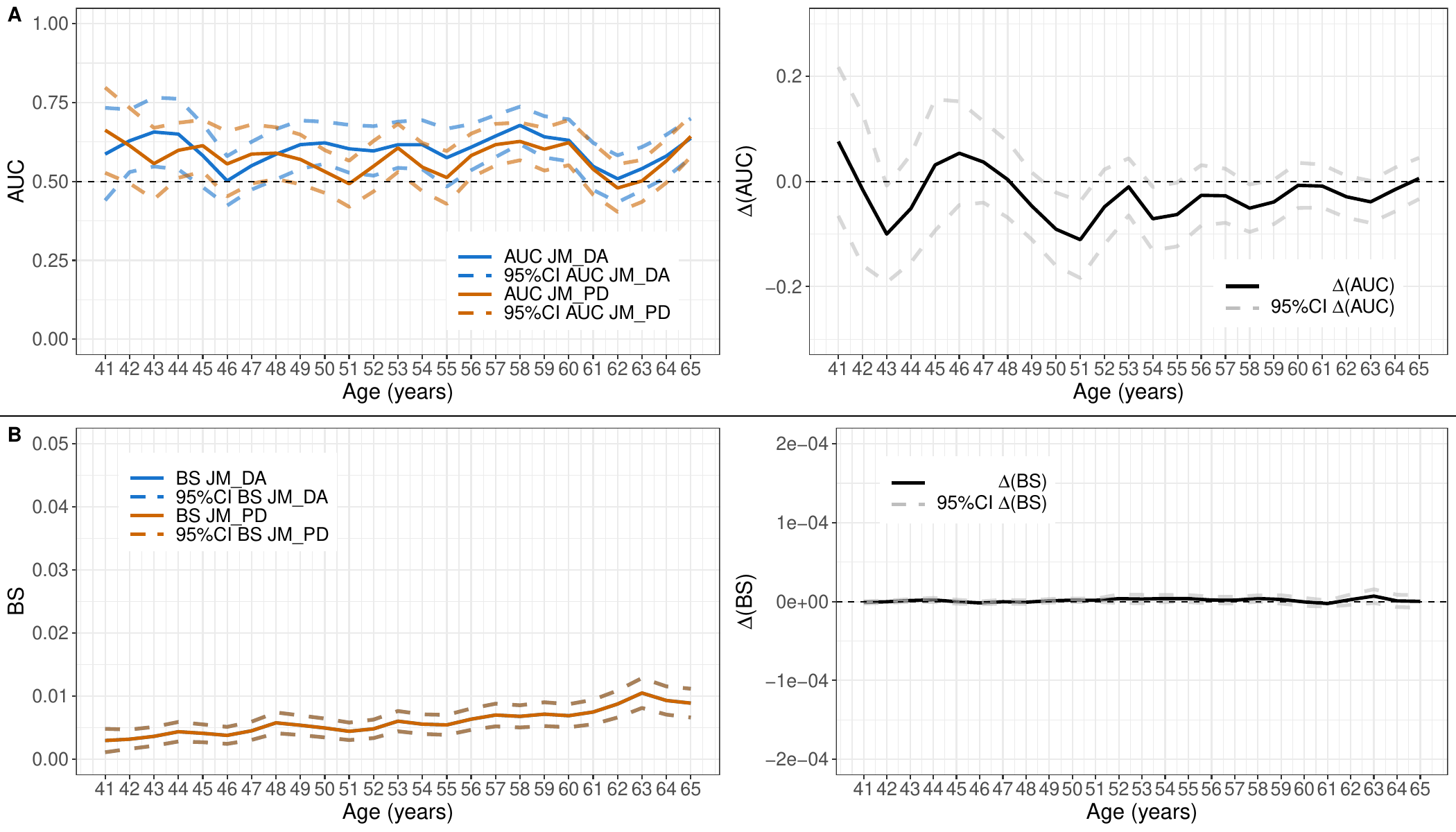}}
\caption{\textbf{Predictive accuracy of the optimal joint model (JM) with either dense area (DA) or percent density (PD) as the longitudinal biomarker within time interval $(s, s + w)$ when $s =\{41, 42, ..., 65\}$ and $w = 2$ years.} $\Delta(AUC)$ and $\Delta(BS)$ denote the differences in AUC and BS between JM with PD versus JM with DA. Dashed lines represent $95\%$ point-wise confidence intervals (CIs)}
\label{f:pred_accuracy_w2}
\end{figure}

\clearpage
\begin{table}[htbp]
\caption{\textbf{Visit-level evaluations of dense area and percent density by BI-RADS categories for $N = 77,298$ screened women.} Dense area (in $cm^2$) and percent density (in \%) were estimated using the new deep learning model of the DeepJoint algorithm.}
\label{t:PD_&_DA_BI-RADS}
\begin{center}
\begin{small}
\begin{tabular}{l@{\hspace{2pt}}c@{\hspace{4pt}}c@{\hspace{4pt}}c@{\hspace{4pt}}c}
\hline
& \multicolumn{4}{c}{\textbf{BI-RADS score}}\\
 & \textbf{A} & \textbf{B} & \textbf{C} & \textbf{D} \\
\textbf{At baseline} & \textbf{N = 9,034} & \textbf{N = 27,988} & \textbf{N = 32,753} & \textbf{N = 7,523} \\
\hline
\textbf{Dense area $(cm^{2})$} & & & & \\
 \ Cancer-free & 19.3 [10.1--35.3] & 59.0 [35.5--89.4] & 98.8 [72.2--131.6] & 114.2 [86.4--151.0] \\
 \ Cancer & 21.0 [14.5--38.6] & 65.2 [41.9--98.5] & 108.1 [84.2--141.0] & 122.6 [98.7--150.5] \\
\hline
\textbf{Percent density $(\%)$} & & & & \\
 \ Cancer-free & 4.9 [2.5--10.0] & 21.4 [11.4--33.7] & 44.5 [34.3--54.3] & 62.3 [53.2--71.6]\\
 \ Cancer & 5.0 [4.5--9.5] & 22.4 [12.6--36.3] & 43.6 [33.8--52.7] & 59.3 [51.3--69.4] \\
\hline
\multicolumn{5}{l}{Data are median [IQR]}
\end{tabular}
\end{small}
\end{center}
\end{table}

\end{document}